\documentclass[12pt,a4paper]{article}
\usepackage[T1]{fontenc}
\usepackage[body={175mm,220mm}]{geometry}

\normalsize

\pdfoutput=1

\usepackage[nottoc,numbib]{tocbibind}

\usepackage{epsfig}
\usepackage{enumitem}
\usepackage{comment}
\usepackage{array}
\usepackage{latexsym}
\usepackage{color}
\usepackage{amssymb}
\usepackage{graphicx}
\usepackage{cite}

\usepackage[english]{babel}
\usepackage [applemac] {inputenc}

\usepackage{MnSymbol}
\usepackage{mathtools}

\newcommand{\be}[0]{\begin{equation}}
\newcommand{\ee}[0]{\end{equation}}
\newcommand{\ba}[0]{\begin{eqnarray}}
\newcommand{\ea}[0]{\end{eqnarray}}

\def\bs{\begin{subequations}}
\def\es{\end{subequations}}

\def\thebibliography#1{%
\vskip 0.5cm \centerline{\bf \Large References}
\list{%
[\arabic{enumi}]}{\settowidth\labelwidth{[#1]} \leftmargin\labelwidth \advance\leftmargin\labelsep
\usecounter{enumi}}
\def\newblock{\hskip .11em plus .33em minus .07em}
\sloppy\clubpenalty4000\widowpenalty4000 \sfcode`\.=1000\relax}

\newcommand{\bea}{\begin{eqnarray}}
\newcommand{\eea}{\end{eqnarray}}

\newcommand{\F}{{\cal F}}
\newcommand{\R}{{\cal R}}
\newcommand{\N}{{\cal N}}

\newcommand{\T}{{\cal T}}

\renewcommand{\S}{{\cal S}}

\renewcommand{\b}{\bar}
\newcommand{\h}{\hat}

\renewcommand{\t}{\tilde}

\renewcommand{\O}{{\cal O}}

\newcommand{\abs}{|}

\renewcommand{\and}{\mbox{and}}

\newcommand{\DAL}{\! \scalebox{1.4}{$\square$}}

\renewcommand{\b}{\bar}

\renewcommand{\t}{\tilde}

\def\cd{{\cal D}}

\def\cf{{\cal F}}

\def\ci{{\cal I}}

\def\cm{{\cal M}}
\def\cn{{\cal N}}

\def\cz{{\cal Z}}

\def\nnR{\nonumber\\}

\newcommand{\demi}{\frac{1}{2}}
\def\lbk{\left[}
\def\rbk{\right]}

\DeclareFontFamily{U}{rsfs}{}         
\DeclareFontShape{U}{rsfs}{m}{n}{<5> rsfs5 <6><7> rsfs7          %
  <8><9><10><10.95><12><14.4><17.28><20.74><24.88> rsfs10}{}     %
\DeclareMathAlphabet{\mathfs}{U}{rsfs}{m}{n}                     %
\newcommand{\mfs}[1]{\mathfs {#1}}                               %

\begin{document}

\begin{titlepage}

\begin{centering}

%

\vspace*{15mm}

%
%

{\Large \bf Lecture notes on thermodynamics of ideal string gases \\ $\ \ \ $ \vspace*{-5mm} \\ and  its application in cosmology}

\vspace{5mm}

{\large Lihui Liu}

{\it Institute for Theoretical Physics, KULeuven}

{\it Celestijnenlaan 200D, 3001 Heverlee}

{ \small lihui@itf.fys.kuleuven.be}

\vspace{5mm}

{ \it Based on lectures given at the Eighth Modave Summer School in Mathematical Physics  \\ on 29th and 30th August 2012, Modave, Belgium}


%
%
%
\end{centering}

\vspace{1cm}

\begin{center}

{\Large \bf Abstract}

\end{center}

\noindent In these lecture notes I give a pedagogical introduction to the thermodynamics of ideal string gases. The computation of thermodynamic quantities in the canonical ensemble formalism will be shown in detail with explicit examples. Attention will be given mainly to the thermodynamical consequences of string degrees of freedom, where I will especially address {\em i)} the Hagedorn temperature, a critical temperature above which the canonical ensemble description breaks down, which can be the onset point of some instability of the string gas; {\em ii)} the phase structure arising from compactification, embodied in the moduli-dependence of the Helmholtz free energy, which corrects the tree-level vacuum and can provide mechanism for moduli stabilization. Then I will briefly explain the implementation of string gas thermodynamics in cosmology, showing a simple example which gives rise to a radiation-dominated early universe. Further phenomenological issues and open questions will be discussed qualitatively with references indicated, including the Hagedorn instability in the resolution of the initial singularity, moduli stabilization, generation of hierarchy, radiative symmetry breaking and primordial cosmological fluctuations.

%
%
%
%
%
%
%
%
%
%
%
%

\end{titlepage}

\tableofcontents

\newpage

\section{Introduction and summary} \label{Intro}

%

The pursuit for the resolution of the initial cosmological singularity has long been one of the major driving forces for research in high energy physics. The common belief is that describing the initial high energy state of the universe is out of the scope of general relativity, and some quantum gravity theory should exist, which can properly describe the spacetime under extreme conditions, and whose low energy limit contains both the general relativity and the standard model.


%




When in the early 1980s physicists became the more and more aware that superstring theory could probably be that very theory of quantum gravity \cite{StringDebut}, attempts also started to investigate string theory effects in early cosmology \cite{Alvarez:1984ee}. The hope was that the Big Bang picture might be radically corrected by the non-local degrees of freedom of strings activated in the high temperature primordial universe, which are not available in quantum field theory. This suggested that the starting point for working out a string resolution of the initial cosmological singularity, could be the thermodynamics of string gases, and indeed a considerable amount of work has been motivated since then on string statistical mechanics \cite{Sundborg:1984uk,AlvarezStFT,Bowick:1985az,Tye:1985jv}.


Being closely related to early cosmology, the high temperature behavior of ideal string gases has been extensively studied. The aspect which received most attention is the presence of a critical temperature, referred to as the {\em Hagedorn temperature} \cite{Hagedorn:1965st}, above which the canonical partition function becomes ill defined. 
In some models the Hagedorn temperature is shown to be the maximum temperature of the string gas, while in others it is shown to lead to a phase transition \cite{Sundborg:1984uk,Bowick:1985az}, which was later realized to be related to the condensation of some thermal winding modes \cite{Atick:1988si,Barbon:2001di,Antoniadis:1991kh,Antoniadis:1999gz}.  
Therefore clarifying the behavior of string gases near this critical temperature, and working out its cosmological implications are of particular interest. However this turned out to be a very tough task due to the lack of analytic control. Although considerable progress has been made, no general conclusive result has been obtained so far. 

In the pioneering work \cite{Brandenberger:1988aj}, an intuitive proposal was made for resolving the initial cosmological singularity, where the mechanism involved exploits the Hagedorn temperature as the maximum temperature of the string gas, as well as the T-duality symmetry which implies that the string gas cannot distinguish large scales from small. A broad range of work have been inspired for elaborating and developing the ideas in \cite{Brandenberger:1988aj}, and this gave birth to the scenario of string gas cosmology (cf. \cite{Brandenberger:2008nx} and the references therein). By considering an ideal string gas coupled to a classical background of dilaton gravity, the string gas cosmology scenario managed to account for the dimensionality of the universe, the stabilization of moduli, and later it was shown to be able to produce a nearly scale-invariant spectrum of primordial cosmological perturbation in a bouncing universe, leading to an alternative scenario to inflation. 

More recently, there is another line of work developed in \cite{Kounnas:2007hb,cosmoB,cosmoC,stabmod1,stabmod2,cosmoreview,Florakis:2010is,Kounnas:2011fk,Kounnas:2011gz,Estes:2011iw,Liu:2011nw}, which attempts to construct a first-principles string theory scenario for cosmology, simply based on string theory subjected to cosmological principles. It turns out that this approach gives rise to a universe filled with an ideal string gas, whose thermal effects can be naturally taken into account. The important phenomenologies such as moduli stabilization, supersymmetry breaking, generation of hierarchy, can be addressed on a solid string theory basis, and
most interestingly, analytic non-singular cosmological solutions are found for certain models \cite{Florakis:2010is,Kounnas:2011fk,Kounnas:2011gz}.



 

These lecture notes are meant to provide an introductory course on ideal string gas thermodynamics, and will also cover, to some extent in the end, its cosmological application figured out in \cite{cosmoB,cosmoC,stabmod1,stabmod2,cosmoreview,Estes:2011iw,Liu:2011nw,Florakis:2010is,Kounnas:2011fk,Kounnas:2011gz}. 
To this end it will be convenient to address the string gas thermodynamics with canonical ensemble formalism, and thus most technical computations to be carried out will be centered around the canonical partition function $\cz={\rm Tr}\, e^{-\beta H}$. At this point we need to illustrate the idea behind the evaluation of $\cz$ for ideal string gases, which provides the line of logic that the lecture notes will follow.

For a generic quantum system at finite temperature, we can perform the Euclidean path integral in the second quantized theory to compute its canonical partition function
\begin{align}
	{\rm Tr}\, e^{-\beta H}=\int\! \cd \phi \; e^{-S[\phi]}, \label{PI1}
\end{align}
where we let $\phi$ account collectively for all the degrees of freedom involved and $S[\phi]$ be the Euclidean action. Finite temperature is implemented by the Euclidean time circle $S^1(R_0)$ whose perimeter is the inverse temperature: $2\pi R_0 =\beta={1\over T}$, and along which bosonic (fermonic) fields take periodic (anti-periodic) boundary condition. 

When dealing with a gas of weakly interacting particles, the path integral can be schematically represented and approximated as
\begin{align}
	\Big({\rm Tr}\, e^{-\beta H} \Big)_{\rm particle} = \exp \bigg[\, \begin{minipage}{10mm} \includegraphics[width=10mm]{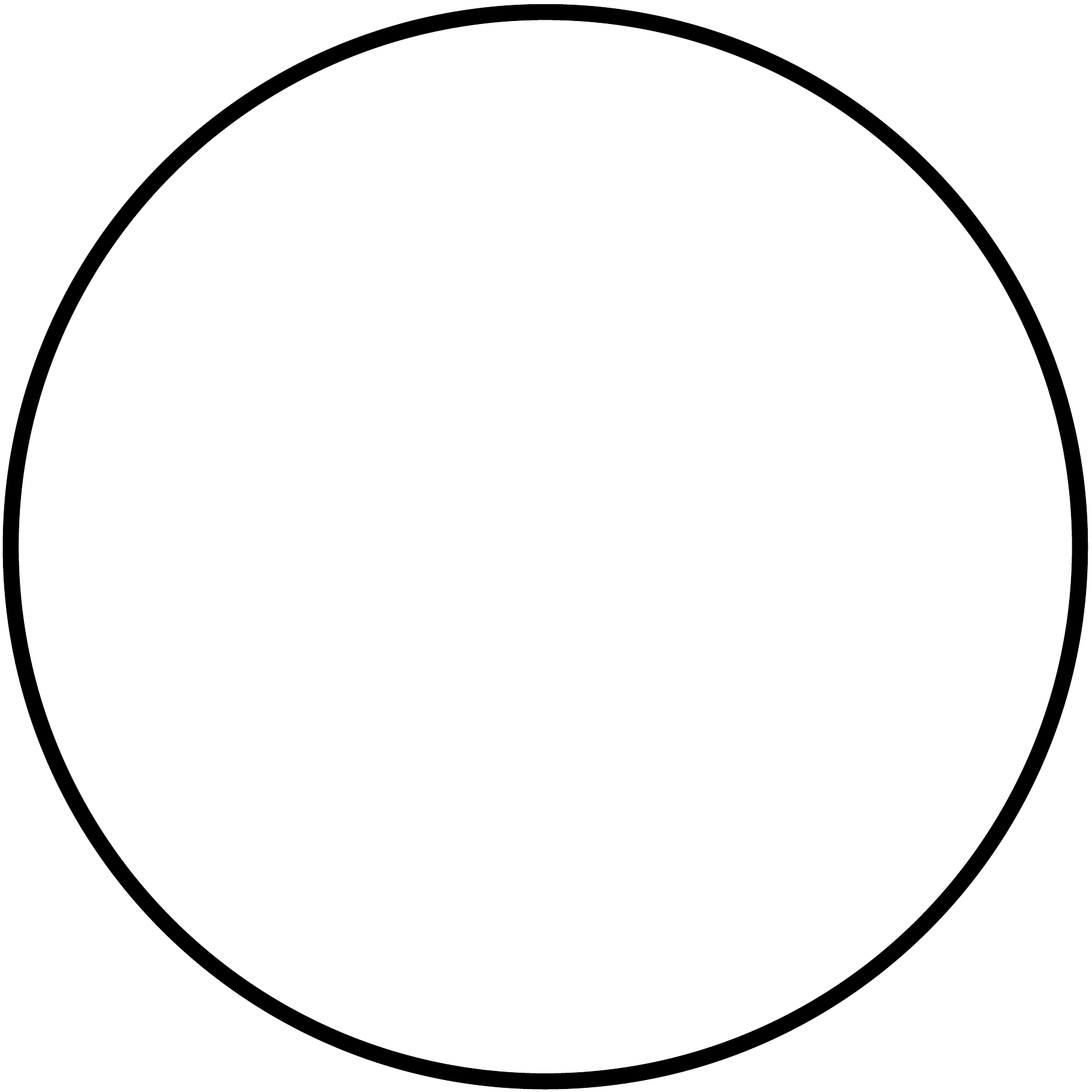} \end{minipage} +\begin{minipage}{15mm} \includegraphics[width=15mm]{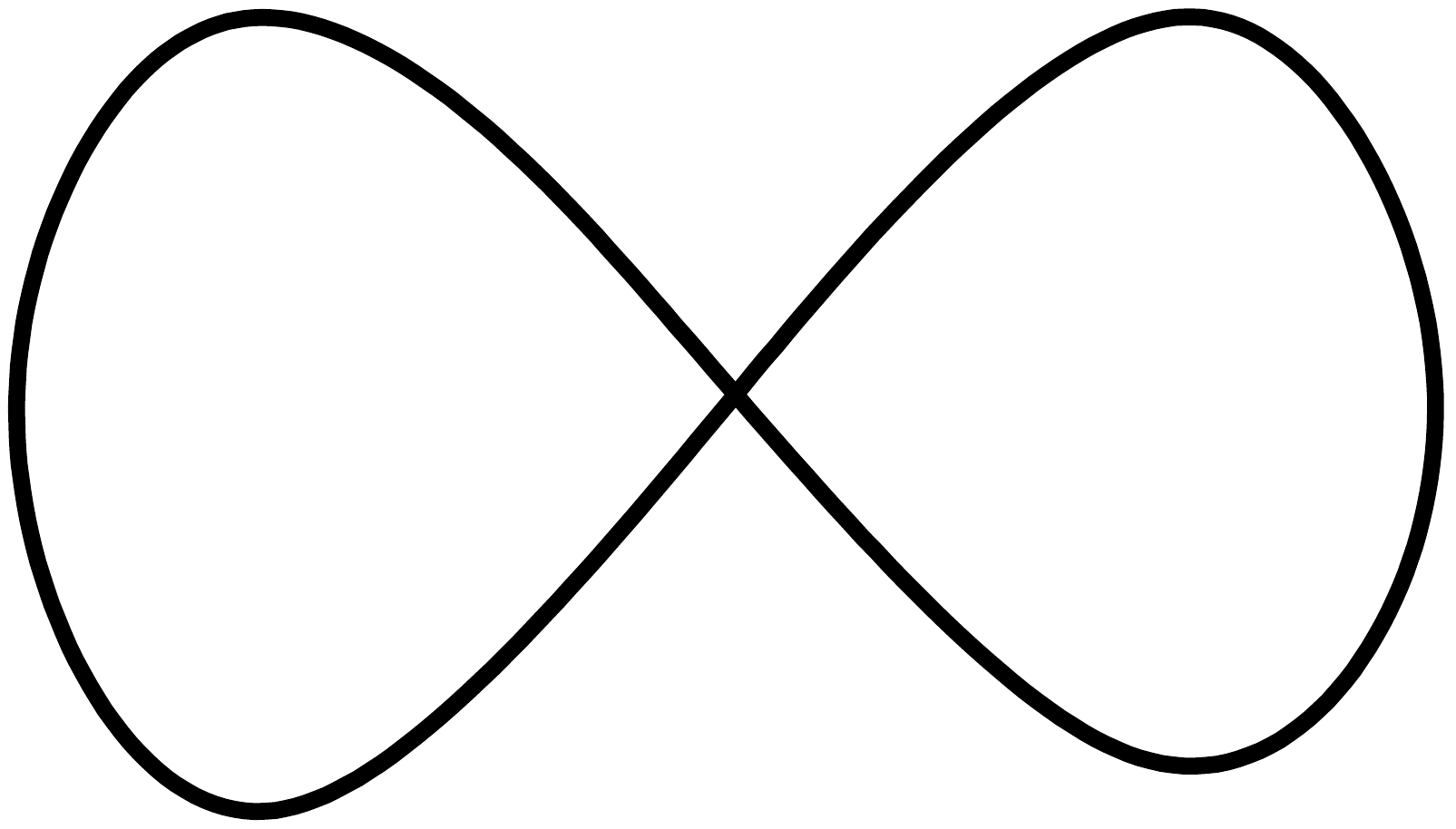} \end{minipage}+\begin{minipage}{10mm} \includegraphics[width=10mm]{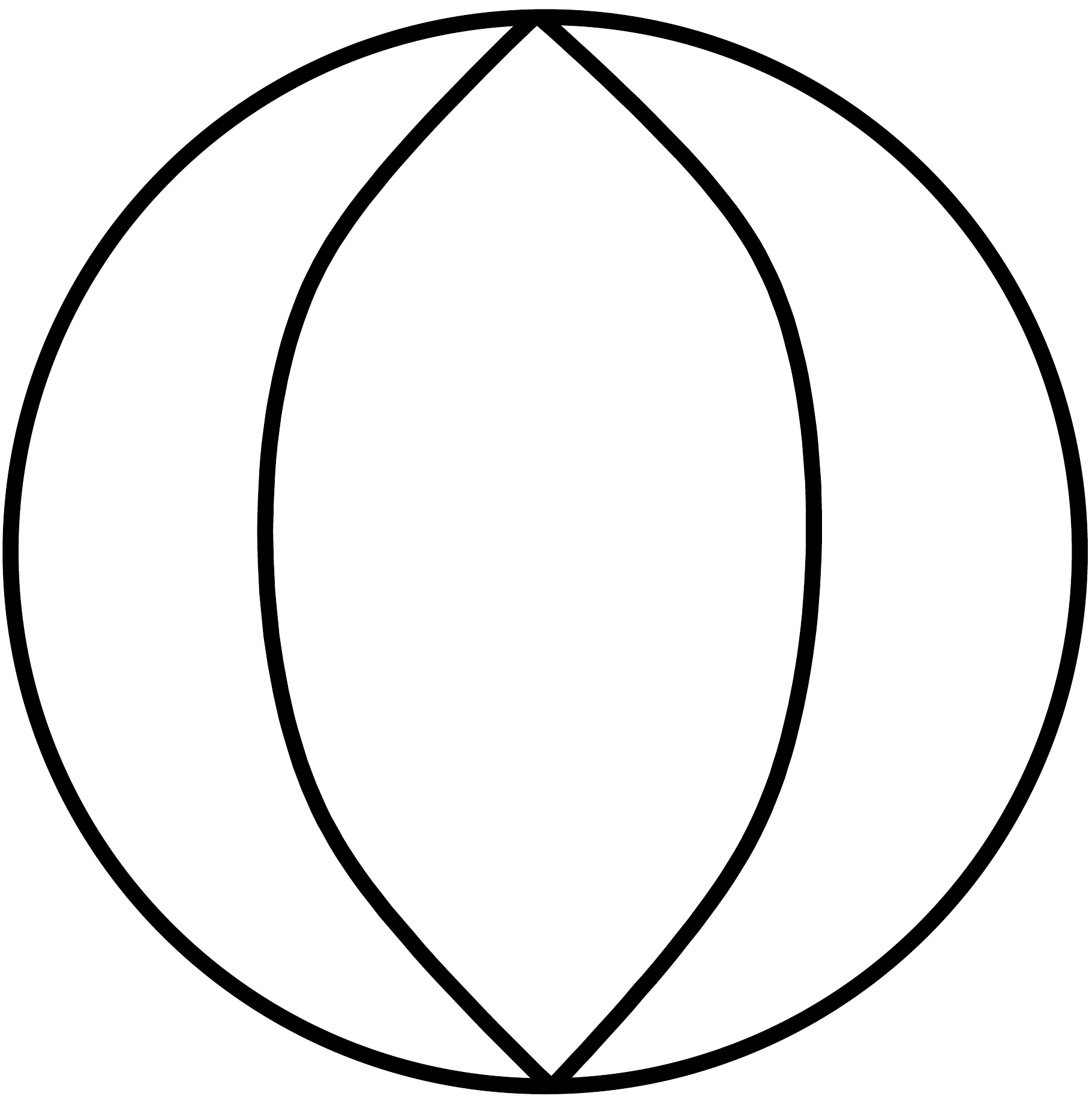} \end{minipage}+ \dots \bigg]\approx  \exp\bigg[\begin{minipage}{10mm} \includegraphics[width=10mm]{circle_no.pdf} \end{minipage}\bigg],\label{Ct1}
\end{align}
which points out that an alternative way to evaluate Eq.(\ref{PI1}) for this gas is to compute the one-loop amplitude $\begin{minipage}{7mm} \includegraphics[width=7mm]{circle_no.pdf} \end{minipage}\,$ in the first quantized theory. We stress that here this one-loop amplitude should be computed against a thermal background with compact Euclidean time $S^1(R_0)$. 
In the same spirit for an ideal gas of strings, which is a quantum system containing infinite degrees of freedom, we can still stick to the methodology Eq.(\ref{PI1}). The difficulty of direct computation, however, is in figuring out the action $S[\phi]$ for second-quantized string theories, which is the concern of string field theory. We will not confront this challenge directly, but will take the alternative path inspired by the particle gas diagrams (\ref{Ct1}), whose string theory version is
\begin{align}
	\Big({\rm Tr}\, e^{-\beta H}\Big)_{\rm string} = \exp\Big[\begin{minipage}{12mm} \includegraphics[width=12mm]{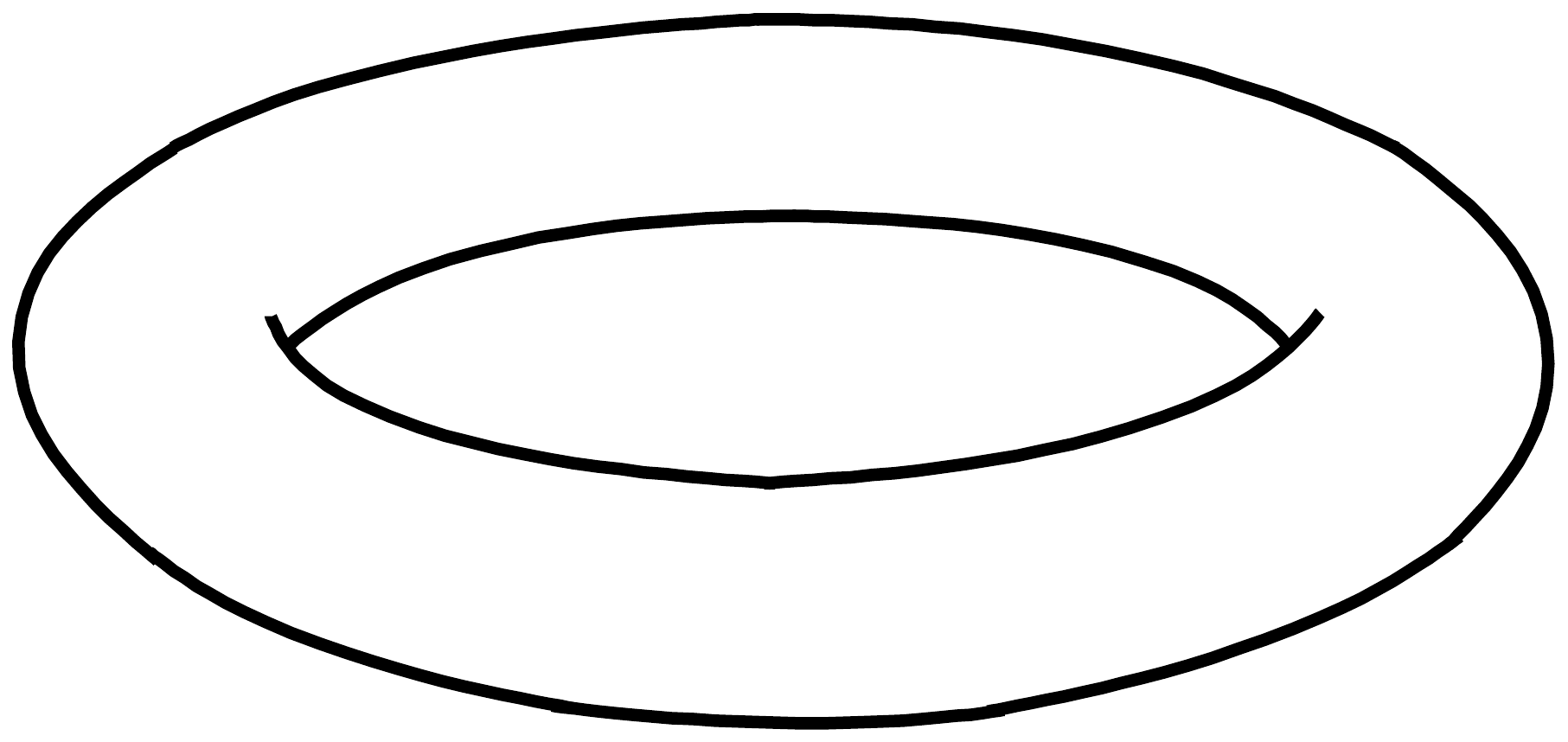} \end{minipage}+ \begin{minipage}{19mm} \includegraphics[width=19mm]{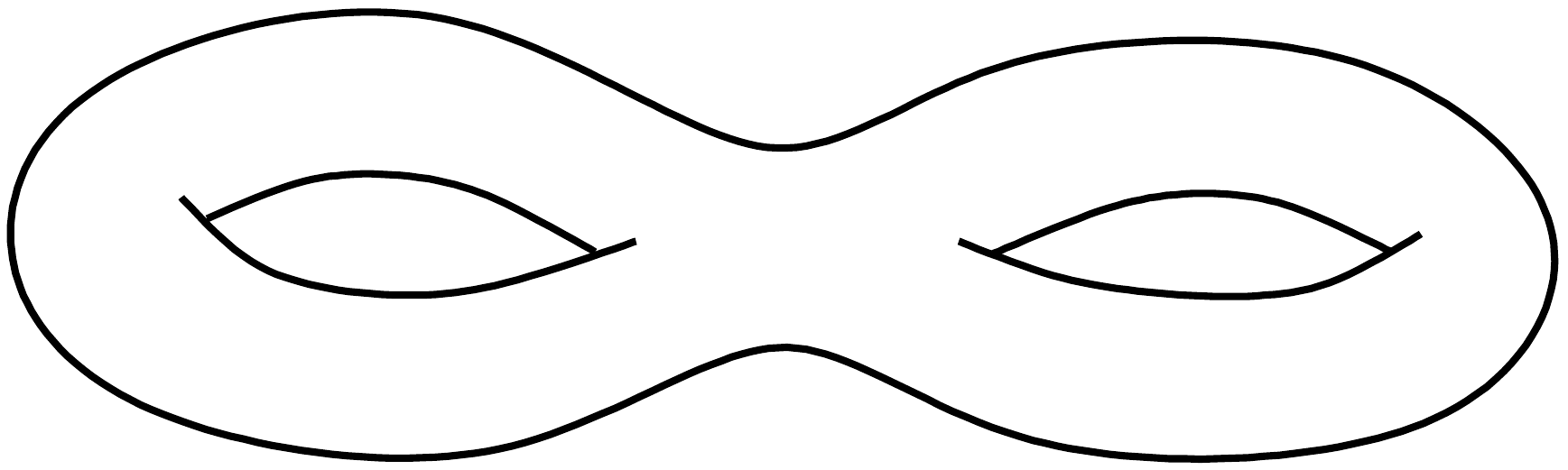} \end{minipage}+\dots \Big]\approx  \exp \Big[\begin{minipage}{12mm} \includegraphics[width=12mm]{torus.pdf} \end{minipage} \Big]. \label{eq3eq3}
\end{align}
Therefore the problem is converted to the computation of the string one-loop (genus-one) vacuum-to-vacuum amplitude $\begin{minipage}{10mm} \includegraphics[width=10mm]{torus.pdf} \end{minipage}$ against a thermal background with compact Euclidean time $S^1(R_0)$. In fact on the practical level the result of $\begin{minipage}{10mm} \includegraphics[width=10mm]{torus.pdf} \end{minipage}$ is more useful than the partition function itself since generally it is $\ln( {\rm Tr}\, e^{-\beta H})$ instead of ${\rm Tr}\, e^{-\beta H}$ that is involved directly in the formulae for computing thermodynamic quantities such as the energy, the entropy etc. Especially we have by definition $ \begin{minipage}{10mm} \includegraphics[width=10mm]{torus.pdf} \end{minipage} = \ln( {\rm Tr}\, e^{-\beta H}) =-\beta F$, where $F$ is the Helmholtz free energy of the ideal string gas. 

For the sake of clarity and explicitness, we will focus our attention on the models where exact computation of $\begin{minipage}{10mm} \includegraphics[width=10mm]{torus.pdf} \end{minipage}$ is achievable, which are basically weakly coupled string theories living in flat spacetime, with toroidal or orbifold compactifications. We will see for such cases that the genus-one amplitudes against a thermal background can be most conveniently obtained by deforming those against a Minkowski background, which we will refer to as {\em thermal deformation}. Although the restriction that we adopt is somewhat severe, the various possibilities of compactifications still leave abundant room for crafting models containing rich phenomenology. 

	
Nevertheless it should be mentioned that some particular curved spacetime backgrounds do allow string thermodynamics to be exactly formulated, and/or be studied in a holographic setting, for example the thermodynamics of little string theory \cite{LSTT}, and strings in the pp-wave background \cite{PandoZayas:2002hh} (c.f. also \cite{Harmark:2006ta} and the references therein). However these topics are beyond the scope of our lecture notes which are oriented rather to cosmological applications.


%

%

The lecture notes will be organized as follows:

In Sec.\ref{NonThAmpSec} we will give a pragmatic account of string one-loop vacuum-to-vacuum amplitudes at zero temperature, providing the preliminary background for the thermal one-loop amplitudes. For simplicity, explicit expressions will be presented only for closed string models, including the bosonic string, the type II and heterotic strings. We will however not display all technical steps, which are some standard textbook computation, but will mainly explain the interpretation of the results in terms of string spectrum, which is helpful for understanding the physics of the thermalization. 

In Sec.\ref{TISG} we will switch on finite temperature to the models covered in Sec.\ref{NonThAmpSec} and compute the resulting thermal one-loop amplitudes, as well as further thermodynamical quantities. Two specific cases, heterotic string gas in 10 dimensions and 9 dimensions, will be investigated respectively in Sec.\ref{10dim} and Sec.\ref{9dim}, by which we aim to unravel their stringy specificities which are also shared by other types of string gases, and which are relevant to the cosmological application. The 10 dimensional example has emphasis on the high temperature behavior where we show especially how the Hagedorn temperature emerges. With the 9 dimensional example we show the phase structure arising from compactification, where the free energy depends nontrivially on the size of the internal space.

In Sec.\ref{GDisc} we will carry out further discussions on high temperature behavior where current understandings of the Hagedorn temperature will be gathered and presented. We will emphasize the following points roughly covered by the three subsections: {\em i)} the Hagedorn temperature is generically present in different types of ideal string gases; {\em ii)} the breaking down of the canonical ensemble formalism can be attributed to some winding modes along the Euclidean time circle becoming tachyonic, the condensation of which can lead to a phase transition; {\em iii)} clarifying the details of the Hagedorn phase is not only significant to the early cosmology, but can also bring along insight into the fundamental degrees of freedom of string theory. 

Finally Sec.\ref{COSapp} implements the explicit results of string gas thermodynamics in Sec.\ref{TISG} in a cosmological context. We will first set up the scenario where we will see that string thermodynamics can naturally arise. In the second part we will deal with a simple example, showing how it leads to a radiation-dominated early universe, which is of course just a tiny step in the long pursuit of a realistic model. Qualitative discussion on further topics and open questions will be presented in the end, including the Hagedorn phase of the universe, moduli stabilization, supersymmetry breaking and hierarchy problem, radiative electroweak symmetry breaking and matter formation, and the primordial cosmological fluctuation.


For notational simplicity, we will use $Z(\beta)$, instead of $\begin{minipage}{7mm} \includegraphics[width=7mm]{circle_no.pdf} \end{minipage}$ or $\begin{minipage}{10mm} \includegraphics[width=10mm]{torus.pdf} \end{minipage}$, to denote $\ln\! \left({\rm Tr} e^{-\beta H} \right)$. We will call $Z(\beta)$ the {\em thermal one-loop amplitude}. We will also need the one-loop amplitude at zero temperature, which we will denote by $Z$, and it will be called the {\em non-thermal one-loop amplitude}. Also the string length will be set to be unit: $l_s=\sqrt {\alpha'}=1$.

\section{Non-thermal one-loop vacuum-to-vacuum amplitudes} \label{NonThAmpSec}

In this section we go through a quick survey on non-thermal string one-loop amplitudes, as a preliminary step towards the thermal one-loop amplitudes in the next section. 
We will mainly focus our attention on the physical idea and especially to the spectral interpretation of the expressions. We will first introduce formal expressions, and then give explicit examples of closed string models in the second part.

\subsection{Formal expressions and their spectral interpretation} \label{sec21}

Here we adopt a heuristic approach to the formulation of non-thermal string one-loop amplitudes, which resorts to the analogy with one-loop amplitudes of particles. Explicit formulae will be shown only for closed strings, while we will be very qualitative with open strings to avoid lengthy technical constructions.

\subsubsection*{From particles to strings}

In the Appendix \ref{app1} the one-loop amplitude of a generic ideal particle gas has been derived. Here we need the result of the non-thermal case Eq.(\ref{lnZtotT0}), which states that for a point particle gas in $d$-dimensional Minkowski spacetime of regularized volume $V_d$, with quantum states of masses $\{M_s\}$ and helicities $\{j_s\}$ ($s$ being the label of states), its non-thermal one-loop amplitude is
\begin{align}
	Z=\sum_{s} {V_{d}\over 2(2\pi)^d }\int_0^{\infty} {d\ell \over \ell^{1+{d\over 2}}}\, (-1)^{2j_s} \exp \big(\!-\pi\, M^2_s\, \ell \big).\label{1AmpG}
\end{align}
Here $\ell$ is the Schwinger time parameter, which tells the proper time that a particle spends for finishing the loop. In order to better establish the analogy between Eq.(\ref{1AmpG}) and its string theory counterpart, we introduce the notation $q:=e^{-2\pi \ell}$, and rewrite Eq.(\ref{1AmpG}) as
\begin{align}
	Z=\sum_{s} {V_{d}\over 2(2\pi)^d }\int_0^{\infty} {d\ell \over \ell^{1+{d\over 2}}} \, (-1)^{2j_s}q^{\, {M^2_s \over 2} }\, . \label{1AmpG2}
\end{align}
This shows that to obtain the non-thermal one-loop amplitude, we can coin up a power series of $q$, whose powers are the mass-sqared over 2 with the masses from the mass spectrum, and whose coefficients are $1$ or $-1$ for bosonic or fermionic states, and when this is done, we integrate over the Schwinger parameter with weight $\ell^{-1-d/2}$ to add up the contribution from the loops of all different sizes.  


The non-thermal string one-loop amplitudes can loosely be regarded as constructed in the same spirit: we simply replace in Eq.(\ref{1AmpG2}) the point particle ``$\, q^{\, {M_s^2\over 2}}\, $'' by string ``$\, q^{\, {M_s^2 \over 2}}\, $'', where now the sum over $s$ runs over the whole string mass tower. There is however important difference between closed strings and open strings. The closed string mass spectrum separates into the holomorphic sector $\{M_{\rm L}\}$ and the anti-holomorphic sector $\{M_{\rm R}\}$, with the physical masses those satisfying the level-matching condition $M_{\rm L}^2 =M_{\rm R}^2\,$;\footnote{In the lightcone gauge, $M_{\rm L}^2$ and $M_{\rm R}^2$ are obtained as the eigenvalues of the lightcone Hamiltonians $L_0$ and $\b L_0$. The level matching condition $M_{\rm L}^2=M_{\rm R}^2$ is due to the constraint $L_0=\b L_0$, which follows from the circle isometry of the closed string worldsheet.} for the open string mass spectrum the level matching condition is automatically satisfied due to the open string boundary conditions, so that there is no need to distinguish the holomorphic masses from anti-holomorphic masses. We now discuss these two cases separately.

\subsubsection*{Closed strings}

Assuming the mass spectra of the holomorphic and the anti-holomorphic sectors of closed strings are $\{M_{\rm L}\}$ and $\{M_{\rm R}\}$, mimicking Eq.(\ref{1AmpG}), we expect the one-loop amplitudes to take the following form
\begin{align}
	\sum_{\rm L, R} {V_{d}\over 2(2\pi)^d } \, (-1)^{2j({\rm L,R})} \int_{-\demi}^{\demi}dv\, \exp\left[\! -\pi (M^2_{\rm L} -M^2_{\rm R}) {v\over 2} \right] \int_0^{\infty} {d\ell \over \ell^{1+{d\over 2}}} \exp\left[\! -\pi (M^2_{\rm L}+M^2_{\rm R}) {\ell \over 2} \right], \label{1AmpS}
\end{align}
where the sum over $\rm L$ and $\rm R$ is understood to run over the holomorphic and the anti-holomorphic sector separately; $j({\rm L,R})$ is the helicity of the state $({\rm L,R})$ in spacetime, for example in type II strings $j({\rm L,R})$ is integer for NS-NS (NS for Neveu-Schwarz) or RR (R for Ramond) states, half-integer for NS-R or R-NS states. The integration over $v$ imposes the level matching condition, so that the sum over $M_{\rm L}$ and $M_{\rm R}$ only picks up $M_{\rm L}^2=M_{\rm R}^2$ terms, the contribution from physical states. The spacetime dimension is left general, denoted by $d$, in case the model is compactified down to lower dimensions. 

Defining the complex parameter $\tau=\tau_1+i\tau_2:=v+i\ell$, usually referred to as the Teichm\"uler parameter, the above expression is rewritten as
\begin{align}
	\sum_{\rm L,R} {V_{d}\over 2(2\pi)^d } \int_{\sqcup} {d\tau_1d\tau_2\over \tau_2^{1+d/ 2}}\, (-1)^{2j({\rm L,R})} \, q^{\, {M_{\rm L}^2\over 4}}\, \b q^{\, {M_{\rm R}^2\over 4}},  \label{1AmpS2}
\end{align}
where $q=e^{2i\pi \tau}$, $\b q=e^{-2\pi i \b \tau}$, and $\sqcup$ denotes the strip defined by $-\demi <\tau_1 <\demi$ and $\tau_2>0$. The above expression is just one step away from the genuine closed string one-loop amplitudes, because the integral over $\tau$ over-counts physically equivalent states. Actually the integration measure $\tau_2^{-2} d\tau_1 d\tau_2$ as well as the integrand $\tau_2^{1-{d\over 2}}\, \sum q^{\, {M_{\rm L}^2\over 4}}\, \b q^{\, {M_{\rm R}^2\over 4}}$ are invariant under the $SL(2,\mathbb Z)$ group generated by $\tau\rightarrow \tau+1$ and $\tau\rightarrow -\tau^{-1}$, which is the residual symmetry after fixing the worldsheet diffeomorphism invariance. In order to remove this redundancy, we have to replace the integration domain by the fundamental domain of the $SL(2,\mathbb Z)$ group, which is defined, up to an $SL(2,\mathbb Z)$-transform, by $-\demi<\tau_1<\demi$ and $\tau_2>\sqrt{1-\tau_1^2}$. Noticeably, the UV divergence is removed because $\tau_2\rightarrow 0$ is excluded from the integration domain. We denote this fundamental domain by $\mfs F$, and hence a generic non-thermal one-loop amplitude of closed strings takes the form
\begin{align}
	Z= \sum_{\rm L,R} {V_{d}\over 2(2\pi)^d } \int_{\mfs F} {d\tau_1d\tau_2\over \tau_2^{1+d/2 }}\, (-1)^{2j(\rm {L,R})} \, q^{\, {M_{\rm L}^2\over 4}}\, \b q^{\, {M_{\rm R}^2\over 4}}.  \label{1AmpS3}
\end{align}
We have several comments at this point

\noindent $\bullet$ The integral domain $\mfs F$ can be rigorously derived from path integral calculation (c.f. \cite{Polchinski:1998rq} Chapter 7, \cite{Lust:1989tj} Chapter 6, or \cite{Polchinski:1985zf} for details). 

\noindent $\bullet$ If one is provided with the one-loop amplitude without knowing the mass spectrum, which is the case in the rest of the lecture notes, one can expand the integrand in a double power series in $q$ and $\b q$, and read off the masses from the powers, degeneracies from the coefficients and tell whether a state is bosonic or fermionic from the sign of the coefficients.

\noindent $\bullet$ The ``Schwinger parameter''  $\tau$, which is in analogy to $\ell$ in the particle case Eq.(\ref{1AmpG2}), is two dimensional. It is because the one-loop diagram, the torus, has two independent deformations which cannot be cancelled by worldsheet diffeomorphism, i.e. the relative size of the two independent non-contractable loops, characterized by $\tau_2$, and the twist when connecting the two ends of the cylinder to form a torus, characterized by $\tau_1$. It is $\tau_2$ which is the analogue of the Schwinger time $\ell$ in Eq.(\ref{1AmpG}).


\subsubsection*{Open strings}

We will not explicitly formulate the open string one-loop amplitudes, since the procedure is technically more involved, while the physical idea remains quite the same as point particles and closed strings. The complexity comes from the problems that open string alone fails to provide massless spacetime states of spin higher than $1$, and it is UV divergent at one-loop level. Two measures are taken to render the model consistent: including a closed string sector, and considering only unoriented string worldsheet. This gives rise to the type I string, whose non-thermal one-loop amplitude is the sum of the amplitudes of four different worldsheet topologies of genus one:
\begin{align}
	 & Z=\, \begin{minipage}{13mm} \includegraphics[width=13mm]{torus.pdf}\end{minipage} \ +\  \begin{minipage}{11mm} \includegraphics[width=11mm]{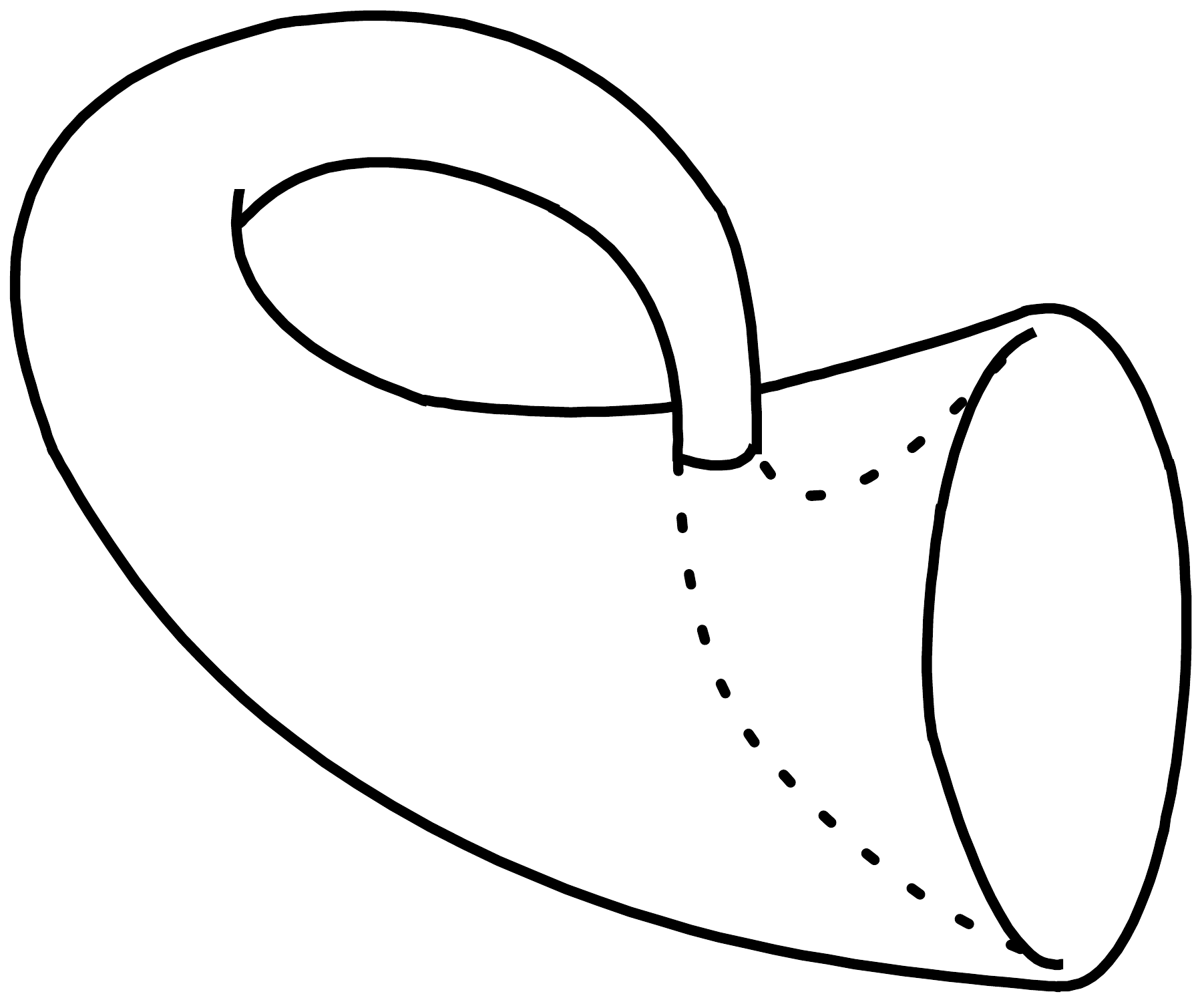}\end{minipage}\ + \ \begin{minipage}{12mm} \includegraphics[width=12mm]{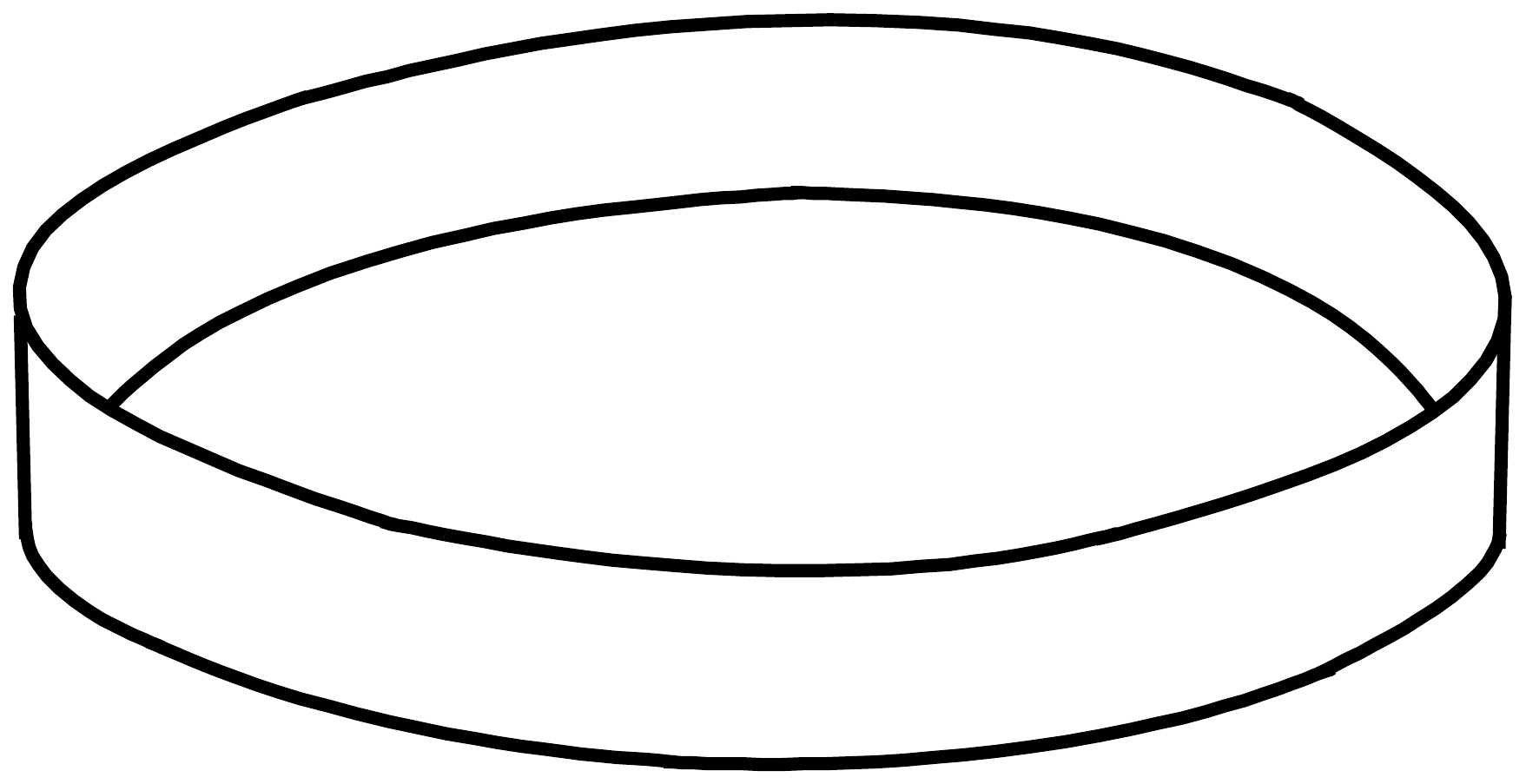}\end{minipage}\ +\ \begin{minipage}{12mm} \includegraphics[width=12mm]{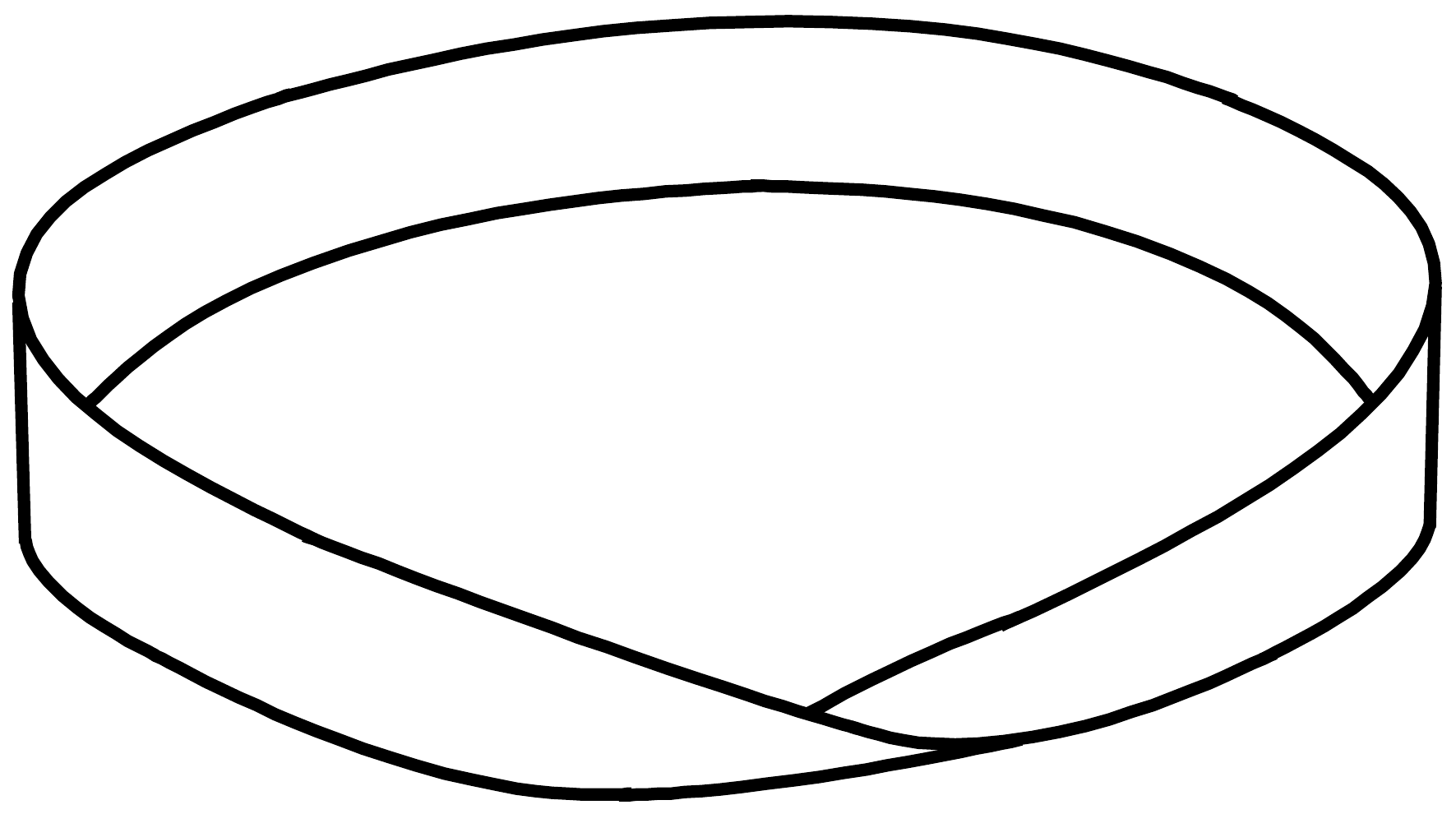}\end{minipage} \ . \label{TotAmpTI} 
\end{align}

\noindent We draw attention to the following points:

\noindent $\bullet$ The spectrum of spacetime states consists of both closed string states and open string states, and all these states are invariant under the reversal of worldsheet orientation (positive worldsheet parity).

\noindent $\bullet$ The torus \begin{minipage}{10mm} \includegraphics[width=10mm]{torus.pdf}\end{minipage} and the Klein bottle \begin{minipage}{9mm} \includegraphics[width=9mm]{KleinBottle.pdf}\end{minipage} amplitudes are from the closed string sector, where the sum of the two projects out the contribution of states of negative worldsheet parity. The same thing is true of the open string sector, where we have the annulus \begin{minipage}{9mm} \includegraphics[width=9mm]{Annulus.pdf}\end{minipage} and the M\"obius strip \begin{minipage}{9mm} \includegraphics[width=9mm]{Moebius.pdf}\end{minipage} amplitudes. 

\noindent $\bullet$ The expression for the torus amplitude is just that of the closed string amplitude that we have described in the previous part; for the other three amplitudes, the expressions are the same as that of the point particle amplitude Eq.(\ref{1AmpG2}), since the geometries of these three types of worldsheet are characterized by only one real parameter which is identified as the Schwinger time parameter.

Detailed construction of consistent open string models and the computation of their one-loop amplitudes can be found in \cite{Angelantonj:2002ct}, or in textbooks \cite{Kiritsis:2007zza,Polchinski:1998rr,Polchinski:1998rq}.

%

\subsection{Some typical examples of closed strings} \label{ExNonTh}

To concretize the discussions in the previous subsection, here we show explicitly the one-loop amplitudes of some typical simple models, where the technicalities involved can be applied to more complicated models. Since the one-loop amplitudes are always an integral over the Teichm\"uler parameter $\tau$ as in Eq.(\ref{1AmpS3}), all we need to show here is how to write down the integrand: the power series in $q$ and $\b q$. It turns out that these power series can be packed into the Dedekind $\eta$-function and the Jacobi elliptic functions or $\theta$-functions, of which the useful notations and definitions are collected in Appendix \ref{tttttt}.

\subsubsection*{Closed bosonic string}

The bosonic string is not a consistent model with its tachyonic ground state, but we still choose it to begin with due to its technical simplicity. The model lives in a 26-dimensional background space, and has 26 worldsheet bosons. Its one-loop amplitude reads
\begin{align}
	Z_{\rm b}={V_{26} \over 2 (2\pi)^{26}} \int_{\mfs F} {d\tau_1 d\tau_2 \over \tau_2^{14}}\, {1 \over \eta^{24} \b \eta^{24}}\, . \label{AmpBos}
\end{align}
The integrand can be conveniently obtained in the lightcone gauge, where we are left with 24 transverse worldsheet bosons, each contributing a factor of $\eta^{-1} \b \eta^{-1}$, accounting exactly for $\eta^{-24} \b \eta^{-24}$. The mass spectrum of spacetime states can be read off from the powers and coefficients in the expansion of the integrand as a power series of $q$ and $\b q$.  More accurately, using the definition (\ref{DefEta}) we have the expansion
\begin{align}
	{1\over \eta^{24} \b \eta^{24}}= q^{-1}\b q^{-1}+24^2 +324^2 \, q\b q+\dots, \label{expBosQQ}
\end{align}
where higher order terms and terms with different powers of $q$ and $\b q$ are omitted. This shows that we have in the physical spectrum, one tachyon of mass-squared $-4\, l_s^{-2}$; $24^2=576$ massless states transforming in $SO(24)_v\times SO(24)_v$ (the subscript $v$ standing for vector representation), containing the dilaton, the graviton and the anti-symmetric tensor; $324^2$ states of mass $2\,l_s^{-1}$, transforming in $SO(25)_{\square\!\square}\times SO(25)_{\square\!\square}$; etc.

\subsubsection*{Type II string}

The type II string lives in a 10-dimensional background space, possessing 10 worldsheet bosons and their superpartner, 10 worldsheet fermions. The one-loop amplitude, in terms of the Dedekind $\eta$-function and Jacobi $\theta$-functions, is
\begin{align}
	Z_{\rm II}= {V_{10} \over 2(2\pi)^{10}}\int_{\mfs F} {d\tau_1\, d\tau_2 \over \tau_2^6}\, {1\over \eta^8 \b\eta^8} \sum_{a,b=0}^1\!\! (-1)^{a+b+ab}\,  {\theta[^a_b]^4\over \eta^{4}}\sum_{\b a, \b b=0}^1\!\! (-1)^{\b a+\b b+\epsilon \b a\b b}\, {\b\theta[^{\b a}_{\b b}]^4\over \b \eta^{4}}, \label{IIABol}
\end{align}
where $\epsilon=0$ or $1$ stand for the type IIA or IIB string respectively. 
We would like to give the following remarks to help understanding this expression: 

\noindent $\bullet$ The integrand can be obtained, as in the previous example of bosonic string, through the lighcone-gauge fixed model. In the holomorphic part, ${\theta[^a_b]^4\over \eta^4}$ arises from the 8 transverse holomorphic worldsheet fermions, the extra $1\over \eta^8$ from the 8 transverse holomorphic worldsheet bosons; the same rule applies to the anti-holomorphic part. The sum over $(a,b)$ and $(\b a, \b b)$ adds up the contribution from all combinations of spin structures. 

\noindent $\bullet$ The indices $a$ and $\b a$, taking independently the values $0$ or $1$, indicate the NS sector or the R sector respectively in the holomorphic and anti-holomorphic sector; the sum over $b$ and $\b b$ implements the GSO projection. In the anti-holomorphic sector, changing the value of $\epsilon$ from $0$ to $1$ reverses the GSO projection in the Ramond sector, which flips the gravitino's chirality, so that the model passes from type IIA to type IIB. 

\noindent $\bullet$ The modular invariance is manifest with Eq.(\ref{IIABol}): using the properties of $\eta$- and $\theta$-functions in the Appendix C of \cite{Kiritsis:2007zza} we can verify the invariance of Eq.(\ref{IIABol}) under $\tau\rightarrow \tau+1$ and $\tau \rightarrow -\tau^{-1}$. 

To let the one-loop amplitude display the structure of the type II string spectrum, one rewrites Eq.(\ref{IIABol}) using the $SO(8)$-characters. Using the definitions Eqs (\ref{SO8chO})--(\ref{SO8chC}), Eq.(\ref{IIABol}) is rewritten as
\begin{align}
	Z_{\rm IIA}= &{V_{10} \over 2(2\pi)^{10}}\int_{\mfs F} {d\tau_1\, d\tau_2 \over \tau_2^6}\, {V_8-S_8 \over \eta^8} {\b V_8-\b C_8 \over \b \eta^8}\nnR = &{V_{10} \over 2(2\pi)^{10}}\int_{\mfs F} {d\tau_1\, d\tau_2 \over \tau_2^6}\, \Bigg[ \bigg({ V_8 \b V_8 \over \eta^8\b \eta^8}+ { S_8 \b C_8 \over \eta^8\b \eta^8} \bigg)- \bigg( {V_8\b C_8  \over \eta^8\b \eta^8} + {S_8\b V_8  \over \eta^8\b \eta^8 }\bigg) \Bigg], \label{IIAol} 
\end{align}
\begin{align}
	Z_{\rm IIB}= &{V_{10} \over 2(2\pi)^{10}}\int_{\mfs F} {d\tau_1\, d\tau_2 \over \tau_2^6}\, {V_8-S_8 \over \eta^8} {\b V_8-\b S_8 \over \b \eta^8} \nnR = &{V_{10} \over 2(2\pi)^{10}}\int_{\mfs F} {d\tau_1\, d\tau_2 \over \tau_2^6}\, \Bigg[ \bigg({ V_8 \b V_8 \over \eta^8\b \eta^8}+ { S_8 \b S_8 \over \eta^8\b \eta^8} \bigg)- \bigg( {V_8\b S_8  \over \eta^8\b \eta^8} + {S_8\b V_8  \over \eta^8\b \eta^8 }\bigg) \Bigg].   \label{IIBol}
\end{align}
In the second line in each of the above two expressions, the integrand is written in the form of [(NS-NS sector)$+$(RR sector)]$-$[(NS-R sector)+(R-NS sector)], where the first half contains bosonic state contributions and the second half fermionic. Numerically Eqs (\ref{IIABol}), (\ref{IIAol}) and (\ref{IIBol}) vanish due to spacetime supersymmetry (c.f. the end of Appendix \ref{tttttt}). However these formal expressions suffice for extracting the spectrum of spacetime states. For example by just staring at Eq.(\ref{IIAol}), we can tell the following information about the massless spectrum : 

\noindent $\diamond$ the NS-NS sector, giving rise to the  $V_8\b V_8$-term, has ground states transforming in $SO(8)_v\times SO(8)_v$, consisting of the dilaton, the graviton and the anti-symmetric tensor; 

\noindent $\diamond$ the RR sector, giving rise to the  $S_8\b C_8$-term, has ground states transforming in $SO(8)_s \times SO(8)_c$ (the subscripts $s$ and $c$ standing for the two spinorial representations of opposite chirality), containing the 1-, 3- and 5-forms; 

\noindent $\diamond$ the NS-R and the R-NS sectors, which correspond to the $V_8\b C_8$-term and the $S_8\b V_8$-term, have ground states transforming in $SO(8)_v\times SO(8)_c$ and $SO(8)_s\times SO(8)_v$, containing the two gravitini of opposite chirality. 

Thus we have seen how to tell the ground states using the formal expressions of one-loop amplitudes, while in order to read off the full physical spectrum, we need to perform the expansion in power series of $q$ and $\b q$, as in Eq.(\ref{expBosQQ}), separately to the four terms in the second line of Eqs (\ref{IIAol}) and (\ref{IIBol}). The same analysis can be applied to the type IIB one-loop amplitude Eq.(\ref{IIBol}), which we do not repeat here. 



Finally we point out that for type II strings, $a+\b a$ plays the role of spacetime fermion number. That is, the spacetime bosons are associated to $a+\b a=0 \, {\rm mod}\, 2$ while spacetime fermions $a+\b a=1\, {\rm mod}\, 2$. Indeed in the second line of Eqs (\ref{IIAol}) and (\ref{IIBol}), the four terms correspond respectively to the values $0$, $2$, $1$, $1$ of $a+\b a$. Thus $2j({\rm L,R})$ in the general expression Eq.(\ref{1AmpS3}) can be concretely written as $a+\b a$.
 
\subsubsection*{Toroidal compactification}

Upon toroidal compactification of closed strings, the compact worldsheet scalars are quantized on the complex plane as
\begin{align}
	X^I_{\rm L}(z)=\demi x^I_{\rm L}- {i\over \sqrt 2} p_{\rm L}^I \ln z+({\rm oscillators}),\ \ \ X^I_{\rm R}(\b z)=\demi x^I_{\rm R}- {i\over \sqrt 2} p_{\rm R}^I \ln \b z+({\rm oscillators}), \label{Xcomp}
\end{align}
where L and R stand for ``left-moving'' and ``right-moving'' corresponding to the holomorphic and anti-holomorphic sectors, $I$ is the index of compact directions, and $x^I_{\rm L,R}$ and $p^I_{\rm L,R}$ are the center-of-mass positions and momenta. Generically $x_{\rm L}^I\neq x_{\rm R}^I$ and $p_{\rm L}^I\neq p_{\rm R}^I$, where the center-of-mass momenta take discrete values lying on the Narain lattice points generated by the basis vectors spanning the internal torus. For example compactifying the 9-th direction on a circle $S^1(R_9)$ results in 
\begin{align}
	p^9_{\rm L,R}={m \over R_9}\mp n R_9\, ,  \ \ \ m,n\in \mathbb Z,\label{p9LR}
\end{align}
where $m$ and $n$ are respectively the quantum number of the internal momentum and of winding along the compact circle $S^1(R_9)$.

More generally, toroidal compactifications down to spacetime dimension $d$, with compact coordinates Eq.(\ref{Xcomp}), induce in the full-dimensional one-loop amplitudes the following modifications: 

\noindent {\it i)} the spacetime dimension is replaced by $d$;

\noindent {\it ii)} the integrand of the integration over $\tau$ acquires a factor of Narain lattice sum 
\begin{align}
	 \Gamma=\sum_{p_{\rm L},p_{\rm R}} q^{p_{\rm L}^2\over 4} \b q^{p_{\rm R}^2 \over 4}, \label{GmTo}
\end{align}
where $p_{\rm L,R}^2:=\sum_I p_{\rm L,R}^I p_{\rm L,R}^I$.

We still take the example of compactification on a circle $S^1(R_9)$ in the 9-th direction where the internal momenta are given by Eq.(\ref{p9LR}). Then the bosonic string one-loop amplitude, given by Eq.(\ref{AmpBos}) for full spacetime dimension, now becomes
\begin{align}
	Z_{{\rm b}/S^1(R_9)} &= {V_{25} \over 2 (2\pi)^{25}} \int_{\mfs F} {d\tau_1\, d\tau_2 \over \tau_2^{27/2}} {1\over \eta^{24} \b\eta^{24}} \sum_{m,n\in \mathbb Z} q^{{1\over 4}\left({m \over R_9}-n R_9\right)^2}\b q^{{1\over 4}\left({m \over R_9}+n R_9\right)^2}; \label{bosolcomp}
\end{align}
also for type II strings, the compactification on $S^1(R_9)$ leads to the following one-loop amplitude in 9 spacetime dimensions:
\begin{align}
	Z_{{\rm II}/S^1(R_9)}& = {V_{9} \over 2(2\pi)^{9}}\int_{\mfs F} {d\tau_1\, d\tau_2 \over \tau_2^{11/2}} {1\over \eta^8 \b\eta^8}\nnR
	\times & \sum_{a,b=0}^1\!\! (-1)^{a+b+ab}\,  {\theta[^a_b]^4\over \eta^{4}}\sum_{\b a, \b b=0}^1\!\! (-1)^{\b a+\b b+\epsilon \b a\b b}\, {\b\theta[^{\b a}_{\b b}]^4\over \b \eta^{4}}\sum_{m,n\in \mathbb Z} q^{{1\over 4}\left({m \over R_9}-n R_9\right)^2}\b q^{{1\over 4}\left({m \over R_9}+n R_9\right)^2}, \label{IIABolcomp}
\end{align}
where $\epsilon$ is as in Eq.(\ref{IIABol}). 

Compactifications generically modify the spectra of spacetime states, where the masses of states can be deformed in function of the moduli fields characterizing the compactification. Here when we expand the integrand of Eqs (\ref{bosolcomp}) and (\ref{IIABolcomp}) in power series in $q$ and $\b q$, the powers acquire $R_9$-dependence, showing that the compactification induces spacetime states with $R_9$-dependent masses.

\subsubsection*{Heterotic string}

The heterotic string demands 10-dimensional background space to be consistent. It has 10 worldsheet bosons, whose holomorphic parts are accompanied with superpartners, the 10 holomorphic worldsheet fermions, and it has 16 internal anti-holomorphic worldsheet bosons compactified on Spin(32)$/\mathbb Z_2$ or $E_8\times E_8$ lattice, giving rise to $SO(32)$ or $E_8\times E_8$ gauge symmetry in spacetime. The one-loop amplitude takes the following form:
\begin{align}
	Z_{\rm het}= {V_{10} \over 2(2\pi)^{10}}\int_{\mfs F} {d\tau_1\, d\tau_2 \over \tau_2^6}\, {1\over \eta^8 \b\eta^8} \sum_{a,b=0}^1\!\! (-1)^{a+b+ab}\,  {\theta[^a_b]^4\over \eta^4} {\b \Gamma_{\rm int} \over \b \eta^{16}}\ ,  \ \ \ {\rm where} \label{HETol} \\ \b \Gamma_{\rm int}= \left\{ \begin{array}{ll} \displaystyle \demi \sum_{\alpha,\beta=0}^1 \b \theta[^{\alpha}_{\beta}]^{16}\ , & SO(32) \text{ heterotic string}, \\ \displaystyle  \bigg( \demi \sum_{\alpha, \beta=0}^1 \b \theta[^{\alpha}_{\beta}]^8\bigg)^{\! 2}, & E_8\times E_8 \text{ heterotic string}. \end{array} \right. \nonumber
\end{align}
In terms of the lightcone-gauge fixed model, $1\over \eta^8 \b \eta^8$ arises from the 8 transverse worldsheet bosons, $\theta[^a_b]^4\over \eta^4$ from the 8 transverse holomorphic worldsheet fermions, and $\b \Gamma_{\rm int}\over \b \eta^{16}$ from the 16 internal anti-holomorphic worldsheet bosons, where $\b\Gamma_{\rm int}$ is the sum of the Spin(32)$/\mathbb Z_2$ lattice or the $E_8\times E_8$ lattice, which is a power series of $\b q$. The modular invariance of Eq.(\ref{HETol}) can be easily verified.

We can also use the $SO(8)$-characters to display the structure of the spectrum.  Using Eqs (\ref{SO8chV})--(\ref{SO8chC}), we have
\begin{align}
	Z_{\rm het}= {V_{10} \over 2(2\pi)^{10}}\int_{\mfs F} {d\tau_1\, d\tau_2 \over \tau_2^6}\, {V_8- S_8\over \eta^8 \b\eta^8}{\b \Gamma_{\rm int} \over \b \eta^{16}}= {V_{10} \over 2(2\pi)^{10}}\int_{\mfs F} {d\tau_1\, d\tau_2 \over \tau_2^6}\, \left( {V_8 \over  \eta^8 }{\b \Gamma_{\rm int} \over \b \eta^{24}} -{ S_8\over \eta^8 } {\b \Gamma_{\rm int} \over \b \eta^{24}}\right) \, ,   \label{HETVSol}
\end{align}
where $\b \Gamma_{\rm int}$ is still as defined in Eq.(\ref{HETol}). In the last step, the integrand is written in the form of (contribution of bosonic states)$-$(contribution of fermonic states).  Here the spacetime bosons and spacetime fermions arise respectively from the NS ($a=0$) and R ($a=1$) sector of the holomorphic worldsheet fermions. Thus here $a$ plays the role of spacetime fermion number and therefore $2j({\rm L,R})$ in the general expression in Eq.(\ref{1AmpS3}) can be concretely written as $a$.

Due to spacetime supersymmetry, Eq.(\ref{HETVSol}) is numerically $0$, but still we can read off the spectrum from the formal expression just as for the type II strings under Eq.(\ref{IIBol}). We will postpone this analysis to Sec.\ref{10dim} where Eq.(\ref{HETVSol}) needs to be more carefully investigated. 


\section{Thermal ideal string gases} \label{TISG}

In this section we consider the thermodynamics of ideal string gases.
We will work out, through thermal deformations, the thermal one-loop amplitudes $Z(\beta)$ based on the non-thermal results in the previous section, and by virtue of Eq.(\ref{eq3eq3}), they are the logarithm of the canonical partition function. Then the whole set of canonical ensemble formalism applies and we can derive all other thermodynamic quantities, for example the energy $E=-\,\partial Z(\beta) /\partial \beta$, the entropy $S=\beta E+Z(\beta)$, the Helmholtz free energy $F=-Z(\beta)/\beta$, and the pressure $P=-\partial F/\partial V$.

In the following we will first specify the rules of thermal deformation of the string one-loop amplitudes, and apply them to the models discussed in the previous section. Then we will investigate in detail two examples of heterotic string gas, where we will put emphasis on the Hagedorn temperature, and the phase structure arising from compactification, which are general features for ideal string gases, and are of interest in string cosmology.

\subsection{From non-thermal to thermal one-loop amplitudes} \label{NTHtoTH}


To figure out how one-loop amplitudes are to be deformed by thermal effects, it is instructive to first look into the same situation for point particles, and we thus invoke the calculations presented in the Appendix \ref{app1}. Comparing the non-thermal one-loop amplitude Eq.(\ref{lnZtot}) and the thermal one Eq.(\ref{lnZtotT0}), which we restate here for clarity:
\begin{align}
	Z &= \sum_s  \frac{ V_d }{2(2\pi)^d} \int_0^{\infty} \frac{d\ell}{\ell^{1+d/2}}\, (-1)^{2 j_s} \exp\Big(\! -\pi \ell M_s^2 \Big)\,, \label{eqz} \\ \label{eqzb}
	Z(\beta) &= \sum_s  \frac{\beta V_{d-1}}{2(2\pi)^d} \int_0^{\infty} \frac{d\ell}{\ell^{1+d/2}}\,\sum_{\t m} (-1)^{2 j_s(\t m+1)} \exp\Big(\! -\frac{\pi R_0^2}{\ell }\, \t  m^2-\pi \ell M_s^2 \Big)\, , 
\end{align}
we find that finite temperature is implemented by inserting in the non-thermal one-loop amplitude the following factor
\begin{align}
	\sum_{\t m\in \mathbb Z} (-1)^{2 j_s \t m}\, \exp\left(- { \pi R_0^2 \over \ell }\,\t m^2\right), \label{Winst1}
\end{align}
and replacing the volume of the full spacetime $V_d$ by that of the thermal background space $\beta V_{d-1}$. We qualify this operation as deformation since Eq.(\ref{eqz}) is the $\beta\rightarrow \infty$ limit of Eq.(\ref{eqzb}). For bosonic states ($2j_s$ even) Eq.(\ref{Winst1}) is merely an instanton sum associated to the compact Euclidean time $S^1(R_0)$, where the instanton number $\t m$ is the number that the particle loop winds the Euclidean time circle. For fermionic states ($2j_s$ odd), the instanton sum in Eq.(\ref{Winst1}) is weighted by a factor $(-1)^{\t m}$, which accounts for the Fermi-Dirac statistics. We call the weight $(-1)^{2j_s \t m}$ in Eq.(\ref{Winst1}) {\it the spin-statistics factor}.

We want to work out the thermal string one-loop amplitudes by a similar deformation, i.e. inserting into the non-thermal expressions the string version of Eq.(\ref{Winst1}) which should be a {\em worldsheet instanton sum weighted by a proper spin-statistics factor}. We know that the worldsheet instanton sum arising from the compactification on a circle is (c.f. for example Sec.4.18 of \cite{Kiritsis:2007zza})
\begin{align}
	\sum_{m,n\in \mathbb Z} e^{-{\pi R_0 \over \tau_2}| \t m-\tau n |^2}, \label{WSinst}
\end{align}
where $\t m$ and $n$ are the numbers that the world sheet winds the Euclidean time circle. There are two winding numbers since the woldsheet is a torus containing two independent non-contractable loops both of which can wind the compact Euclidean time. 
We still need to find out the spin-statics factor weighting the instanton sum, and it turns out that its explicit form depends on the type of string theory. 

Below we will present the results for the specific models that we have discussed in Sec.\ref{ExNonTh}, some intuitive reasonings and guessworks are inevitably involved. Rigorous derivation can be found in \cite{Atick:1988si} for heterotic strings, the logic of which can be applied to other types of strings. The formalism that we present here is are those in \cite{Kounnas:1989dk} where the expressions were cast into a concise form. We will present the formal expressions all at once, leaving explicit calculation and interpretations in later subsections.

\subsubsection*{Bosonic string}

To obtain the thermal bosonic string one-loop amplitude, it suffices to simply insert Eq.(\ref{WSinst}) into Eq.(\ref{AmpBos}), because there are only bosons in the spectrum of spacetime states. Therefore the thermal one-loop amplitude in 26 dimensions is
\begin{align}
	Z_{\rm b}(\beta)={\beta V_{25} \over 2 (2\pi)^{26}} \sum_{\t m,n} \int_{\mfs F} {d\tau_1 d\tau_2 \over \tau_2^{14}}\, e^{-{\pi R_0^2 \over \tau_2}|\t m-\tau n |^2} \, {1 \over \eta^{24} \b \eta^{24}}\, . 
	\label{bTHol}
\end{align}
This means that the thermalization of the bosonic string is just a naive toroidal compactification of the Euclidean time. Indeed when applying Poisson resummation Eq.(\ref{form9}) to the sum over $\t m$ above, we obtain exactly the lattice sum (\ref{GmTo}) with the internal momenta given by Eq.(\ref{p9LR}), $R_9$ replaced by $R_0$.

In case of additional toroidal compactification in spatial dimensions, the thermal one-loop amplitude is still obtained through the steps leading to Eqs (\ref{bosolcomp}) and (\ref{IIABolcomp}) regardless of thermalization, that is, replacing in Eq.(\ref{bTHol}) the full dimension 26 by the number of dimension after compactification, and inserting the Narain lattice sum corresponding to the compact space.

\subsubsection*{Type II strings}

For type II superstrings, as the spectrum consists of spacetime bosons and fermions, the spin-statistics factor is nontrivial. We have already seen in Sec.\ref{ExNonTh} that in type II strings it is $a+\b a$ as in Eq.(\ref{IIABol}) that plays the role of $2j_s$ in Eq.(\ref{Winst1}). Therefore the spin-statistics factor contains at least $(-1)^{(a+\b a)\t m}$. However there should be more than that: since the modular transforms $\tau\rightarrow \tau+1$ and $\tau\rightarrow -\tau^{-1}$ mix up $a$ and $b$ in Eqs (\ref{IIABol}), these two indices need to be treated on equal footing, and the same thing is true of $\b a$ and $\b b$. It can be shown that the result is modular invariant if we also include the factor $(-1)^{(b+\b b)n}$. Thus the complete instanton sum for type II strings is
\begin{align}
	\sum_{\t  m,n\in \mathbb Z} (-1)^{(a+\b a) \t m+(b+\b b)n}\, e^{-{\pi R_0^2 \over \tau_2}|\t m-\tau n |^2}, \label{instII}
\end{align}
and the thermal one-loop amplitude is
\begin{align}
	Z_{\rm II}(\beta)= {\beta V_9 \over 2(2\pi)^{10}}\int_{\mfs F} {d\tau_1\, d\tau_2 \over \tau_2^6}\,  \sum_{a,b} \sum_{\b a, \b b} & \left[\sum_{\t m,n} (-1)^{(a+\b a) \t m+(b+\b b)n}\, e^{-{\pi R_0^2 \over \tau_2}|\t m-\tau n |^2}\right]\, \nnR  &\times {1\over \eta^8 \b\eta^8} \, (-1)^{a+b+ab}\,  {\theta[^a_b]^4\over \eta^{4}}\, (-1)^{\b a+\b b+\epsilon \b a\b b}\, {\b\theta[^{\b a}_{\b b}]^4\over \b \eta^{4}}\, , \label{IIABolthm}
\end{align}
where $\epsilon$ is as in Eq.(\ref{IIABol}), and the instanton sum should be inserted inside the sums over spin structures $(a,b)$ and $(\b a,\b b)$. 

It is not surprising that the insertion of Eq.(\ref{instII}) preserves modular invariance, since this deformation is nothing but a {\em Scherk-Schwarz compactification}\footnote{It was first introduced in field theory\cite{Scherk:1978ta} as a mechanism of spontaneous supersymmetry breaking and later on realized in string theory \cite{SSstring,Kounnas:1989dk}.} of the Euclidean time, which is a freely acting orbifold compactification and which by construction preserves modular invariance. The modular invariance at finite temperature was studied as early as \cite{O'Brien:1987pn}, but the connection with Scherk-Schwarz compactification was realized in \cite{Kounnas:1989dk}.

\subsubsection*{Heterotic string}

We have stated by the end of Sec.\ref{NonThAmpSec} that for the heterotic string, the spacetime fermion number is given by the index $a$ in Eq.(\ref{HETol}). Thus by the same consideration as for type II strings, the full instanton sum to be inserted into Eq.(\ref{HETol}) to account for finite temperature is
\begin{align}
	\sum_{\t m,n\in \mathbb Z} (-1)^{a\t m+bn+\t mn}\, e^{-{\pi R_0^2 \over \tau_2}|\t m-\tau n |^2}. \label{instHET}
\end{align}
The thermal one-loop amplitude is
\begin{align}
	\!\! \!\! Z_{\rm het}(\beta)\! =\! {\beta V_9 \over 2(2\pi)^{10}} \!\!\int_{\mfs F}\!\! {d\tau_1\, d\tau_2 \over \tau_2^6}\, \sum_{a,b} \left[ \sum_{\t m,n}(-1)^{a\t m+bn+\t mn}\, e^{-{\pi R_0^2 \over \tau_2}|\t m-\tau n |^2} \! \right] \! {1\over \eta^8 \b\eta^8}\, (-1)^{a+b+ab}\,  {\theta[^a_b]^4\over \eta^4} {\b \Gamma_{\rm int} \over \b \eta^{16}} ,  \label{HETolthm} 
\end{align}
where again, $\b \Gamma_{\rm int}$ is the lattice sum defined in Eq.(\ref{HETol}). Here just as for the type II case, the insertion of Eq.(\ref{instHET}) implements a Scherk-Schwarz compactification of the Euclidean time.

\subsubsection*{Open string in short}

Recall that the non-thermal one-loop amplitude of open string models is the sum of the contributions from four different worldsheet topologies, which is mentioned by the end of Sec.\ref{sec21} in Eq.(\ref{TotAmpTI}). Thus once the model is thermalized, those four amplitudes (torus, Klein bottle, annulus and M\"obius strip) are deformed separately into their thermal version, and they sum up to the total thermal one-loop amplitude. 
The torus amplitude is just the closed string amplitude, and thus its thermal deformation is just as explained in the previous paragraphs. For the three other amplitudes, since they take the form of one-loop amplitudes of particles as is stated below Eq.(\ref{TotAmpTI}), their thermal deformation proceeds like that for particles, yielding expressions of the form Eq.(\ref{eqzb}). 

%

From a technical point of view, the thermal deformation of open string one-loop amplitudes amounts to performing the Scherk-Schwarz compactification separately to the non-thermal torus, Klein bottle, annulus and M\"obius strip amplitudes, and then summing them up. 

More about Scherk-Schwarz compactification in open string models can be found in \cite{Angelantonj:2002ct}. A detailed computation of type I thermal one-loop amplitude is presented in the appendix of \cite{Estes:2011iw}.


\subsection{Example 1: ideal heterotic string gas in 10 dimensions and Hagedorn temperature} \label{10dim}

To gain a flavor of what all the formulae in the last subsection are about, we will study simple cases of weakly coupled heterotic string thermodynamics where we will focus on the stringy properties commonly shared with other types of string gases. Also the technique to be used is of more general applicability.

As a first example we consider heterotic string in 10 dimensions, whose thermal one-loop amplitude is given by Eq.(\ref{HETolthm}). To perform analytic calculation, we can use the unfolding technique summarized in Appendix \ref{app2}, which decomposes the instanton sum of $(\t m,n)$ into the $(0,0)$-orbit, integrated over the fundamental domain $\mfs F$, and the $(\t m,0)$-orbit where $\t m\neq 0$, integrated over the upper strip $-\demi <\tau_1 <\demi$, $\tau_2>0$ denoted by $\sqcup$. The decomposition leads to
\begin{align}
	Z_{\rm het}(\beta) =& {\beta V_{9} \over 2(2\pi)^{10}} \int_{\mfs F} {d\tau_1\, d\tau_2 \over \tau_2^6}\, \sum_{a,b} {1\over \eta^8 \b\eta^8}\, (-1)^{a+b+ab}\,  {\theta[^a_b]^4\over \eta^4} {\b \Gamma_{\rm int} \over \b \eta^{16}}\nnR    &+	{\beta V_9 \over 2(2\pi)^{10}} \int_{\sqcup} {d\tau_1\, d\tau_2 \over \tau_2^6}\, \sum_{a,b} \left[ \sum_{\t m\neq 0}(-1)^{a\t m}\, e^{-{\pi R_0^2 \over \tau_2}\t m^2} \right]{1\over \eta^8 \b\eta^8}\, (-1)^{a+b+ab}\,  {\theta[^a_b]^4\over \eta^4} {\b \Gamma_{\rm int} \over \b \eta^{16}},\label{HETthpre}
\end{align}
or in terms of the $SO(8)$ characters,
\begin{align}
	 Z_{\rm het}(\beta)=& {\beta V_{9} \over 2(2\pi)^{10}} \int_{\mfs F} {d\tau_1\, d\tau_2 \over \tau_2^6}\, {V_8-S_8\over \eta^8}\,  {\b \Gamma_{\rm int} \over \b \eta^{24}}\nnR    &+	{\beta V_9 \over (2\pi)^{10}} \int_{\sqcup} {d\tau_1\, d\tau_2 \over \tau_2^6}\,\sum_{\t m\neq 0} e^{-{\pi R_0^2 \over \tau_2}\t m^2}\, {V_8-(-1)^{\t m} S_8\over 2\, \eta^8 }\, {\b \Gamma_{\rm int} \over \b \eta^{24}} .  \label{HETthDec}  
\end{align}
The first lines of the above two equations are the $(0,0)$-orbit, which is just the non-thermal one-loop amplitude (only with $V_{10}$ replaced by $\beta V_9$), and which vanishes numerically by virtue of the identity (\ref{JID}), consequence of spacetime supersymmetry. The second lines of the above two equations, which vanish when temperature is switched off ($\beta = 2\pi R_0 \rightarrow \infty$), account for thermal effects. In the sum of $\t m$ in the second lines, we notice that only the terms with odd $\t m$ are non-vanishing; terms with even $\t m$ vanish again due to the identity (\ref{JID}). Discarding all the vanishing pieces in Eq.(\ref{HETthDec}), we are left with
\begin{align}
	Z_{\rm het}(\beta) =&	{\beta V_9 \over 2(2\pi)^{10}} \int_{\sqcup} {d\tau_1\, d\tau_2 \over \tau_2^6}\, \sum_{a,b} \left[ \sum_{\t \jmath \in \mathbb Z}\, e^{-{\pi R_0^2 \over \tau_2}(2\t \jmath+1)^2} \right]{1\over \eta^8 \b\eta^8}\, (-1)^{b+ab}\,  {\theta[^a_b]^4\over \eta^4} {\b \Gamma_{\rm int} \over \b \eta^{16}}\nnR =& {\beta V_9 \over (2\pi)^{10}} \int_0^{\infty}{d\tau_2\over \tau_2^6}  \int_{-\demi}^{\demi} d\tau_1\, \sum_{\t \jmath \in \mathbb Z}\, e^{-{\pi R_0^2 \over \tau_2}(2\t \jmath+1)^2}\,   {V_8+S_8\over  2\, \eta^8 } {\b \Gamma_{\rm int} \over \b \eta^{24}}\, .  \label{HETthDec2}  
\end{align}
To further on, we can power expand the integrand separately in the holomorphic and the anti-holomorphic sector in terms of $q$ and $\b q$ (which can be easily done with Mathematica) and this yields\footnote{We take advantage of these expansions to show explicitly how to read off the spectrum of spacetime states, following the recipe given in Sec.\ref{ExNonTh}. What we can learn from Eqs (\ref{Expan1}) and (\ref{Expan2}) is that, in either the bosonic sector or the fermonic sector, the holomorphic mass tower $\{M_{\rm L}\}$ is given by $M_{\rm L}^2=4K$ ($K=0,1,2,\dots$) in the unit of $l_s^{-2}$, with degeneracy $\S_K$ for the level $K$; and the anti-holomorphic mass tower $\{M_{\rm R}\}$ is given by $M_{\rm R}^2=4\b L$ ($\b L=-1,0,1,2,\dots$),  with degeneracy $\T_{\b L}$ for level $\b L$. The anti-holomorphic sector contains a tachyonic state $\b L=-1$, but it is not physical since physical states need to be level-matched: $M_{\rm L}^2=M_{\rm R}^2$, and in the holomorphic sector the level at $K=-1$ does not exist. Thus the tower of physical bosonic or fermionic states has mass spectrum $M_{\rm ph}^2=4K$ ($K=0,1,2,\dots$), with degeneracy $\S_K\T_K$ for level $K$, e.g. we have $\S_0\T_0=16\times 504$ bosonic (or fermionic) states at the massless level, $\S_1 \T_1=256 \times 73764$ bosonic (or fermionic) states of mass $2 l_s^{\, -1}$, etc.}
\begin{align}
	&{V_8+S_8\over  2\, \eta^{12} }=\!\!\sum_{K=0}^{\infty} \S_{K}\, q^K , \ {\rm where}\ \{\S_K \}_{\!{}_{K \geq 0}}=\{16,256,2304,15360,84224,400896,1711104,\dots\}; \label{Expan1}\\
	&{\b \Gamma_{\rm int} \over \b \eta^{24}}=\sum_{\b L =-1}^{\infty} \T_{\b L}\, \b q^{\, \b L},\ \ \ {\rm where}\ \{\T_{\b L}\}_{\!{}_{\b L\geq -1}}=\{1,504,73764,2695040,54755730,778640256,\dots\}. \label{Expan2}
\end{align}
Here the powers of $q$ and $\b q$ are all integers, which is not necessarily the case in general. Now Eq.(\ref{HETthDec2}) becomes
\begin{align}
	Z_{\rm het}(\beta) = {\beta V_9 \over (2\pi)^{10}} \int_0^{\infty}{d\tau_2\over \tau_2^6}  \int_{-\demi}^{\demi} d\tau_1\, \sum_{\t \jmath  \in \mathbb Z}\, e^{-{\pi R_0^2 \over \tau_2}(2\t \jmath +1)^2}\, \sum_{\tiny \substack{K\geq0\\ \b L\geq -1}} \, \S_K \T_{\b L}\,\, q^K \b q^{\, \b L}\, .  \label{HETthDec3}  
\end{align}
The integration over $\tau$ is now easy to perform. Taking into account the definition $q=e^{2\pi i \tau}$ and the fact that $K$ and $\b L$ are integers, we see that integrating over $\tau_1$ eliminates all terms with $K \neq \b L$ in the sum in Eq.(\ref{HETthDec3}). Thus
\begin{align}
	Z_{\rm het}(\beta) = &{\beta V_9 \over (2\pi)^{10}} \int_0^{\infty}{d\tau_2\over \tau_2^6}  \, \sum_{\t \jmath \in \mathbb Z}\, e^{-{\pi R_0^2 \over \tau_2}(2\t \jmath+1)^2}\, \sum_{K \geq0} \, \S_K\T_K\,\, (q\b q)^K \nnR =& {\beta V_9 \over (2\pi)^{10}} \int_0^{\infty}{d\tau_2\over \tau_2^6}  \, \sum_{\t \jmath \in \mathbb Z} \sum_{K \geq0}\,  \S_K \T_K\, e^{-{\pi R_0^2 \over \tau_2}(2\t \jmath+1)^2-\pi\,\tau_2\, (4K)}\nnR =& (\beta V_9) \beta^{-10}  \sum_{K\geq 0} \S_K \T_K \, G_{10}\big(\beta \, 2\sqrt K \big).  \label{HETthDec4}  
\end{align}
From Eq.(\ref{HETthDec3}) to the first line above, we notice that the tachyonic state contribution in the anti-holomorphic sector, the $\b L=-1$ term in the expansion (\ref{Expan2}), is discarded. Therefore in the second line $K$ is non-negative, which ensures the convergence of the $\tau_2$-integration for each single term in the summation over $K$ and $\t \jmath$.\footnote{Had the power expansion in $q$ in the holomorphic sector of Eq.(\ref{Expan1}) had a tachyonic $q^{-1}$ term, which is the case of bosonic string as shown in Eq.(\ref{expBosQQ}), giving rise to a physical tachyonic state, then the sum over $K$ in Eq.(\ref{HETthDec4}) would contain a $K=-1$ term, which makes the integral over $\tau_2$ diverge at the upper limit $\tau_2\rightarrow \infty$. Generically the presence of a tachyonic state in the physical spectrum makes the one-loop amplitude ill defied.} Therefore assuming the integration and the summations are interchangeable, we did the $\tau_2$-integration and obtained the third line above, where we have defined the function
\begin{align}\label{defG}
	G_d(x)=\sum_{ j\in \mathbb Z} 2\left({x \over  2\pi \, \abs 2 j +1\abs }\right)^{\! {d\over 2}} K_{d\over 2}\big(x\,  \left| 2 j+1\right|\big)\,; \ \ {\rm here}\ d=10.
\end{align}
The function $K_{\nu}(\,\cdot\,)$ is the modified Bessel function of the second kind, which arises by virtue of Eq.(\ref{Knu}).
Based on the properties of $K_{\nu}(\,\cdot\,)$ we know that $G_d(x)$ is a monotonously decreasing function for $x>0$ and it has the following asymptotic behaviors for $d\geq 3$:
\begin{align} \label{asyG}
	G_d(x)=c_d-{c_{d-2}\over 4\pi}\, x^2+\O(x^4) \; \;\; \;({\rm for}\ x\simeq 0)\; , \quad G_d(x)\sim 2\left({x\over 2\pi}\right)^{d-1\over 2} e^{-x} \; \;  \; ({\rm for}\ x\gg 1)\, .
\end{align}
where
\begin{align}
\label{DefcD}
	c_d=G_d(0)={\Gamma({d\over 2})\over \pi^{d\over 2}}\sum_{j\in \mathbb Z}{1\over \abs 2 j+1\abs ^d}\; .
\end{align}
The function $G_d(\,\cdot\,)$ can be written in a more familiar form with the standard canonical ensemble formalism. Referring to the formulae (\ref{FMLA8})--(\ref{lnZstand})
we have, for a given temperature $\beta$ and mass $M$,
\begin{align}
	G_d(\beta M)={\beta^{d-1} \over (2\pi)^{d-1}} \int \!\! d\vec p\, \ln\! \left({1+e^{-\beta \sqrt{\vec p^2+M^2}} \over 1- e^{-\beta \sqrt{\vec p^2+M^2}}}\right), \label{GlnSusy}
\end{align} 
where $\vec p$ is the $(d-1)$-dimensional spatial momentum. Inserting this into Eq.(\ref{HETthDec4}) setting $M=2\sqrt K$, we obtain exactly the standard canonical ensemble formalism for a supersymmetric ideal gas.

 
We give some comments to close the investigation on this model, while the points that we make below can be generalized to other models covered in Sec.\ref{ExNonTh}. 

\noindent $\bullet$ This model has spacetime supersymmetry, where the supersymmetric boson-fermion pairs have masses $ 2 \sqrt K$ ($K=0,1,2,\dots$). Each boson-fermion pair of mass $2 \sqrt K$ contribute to the thermal one-loop amplitude the amount $\beta V_{10} \times \beta^{-10}\, G_{9}(\beta\, 2\sqrt K)$. More generally for an ideal string gas having spacetime supersymmetry in dimension $d$, each boson-fermion pair of mass $M_{\rm phy}$ contributes $\beta V_{d-1} \times \beta^{\, -d}\, G_{d}(\beta  M_{\rm phy})$. Referring to the asymptotic behaviors (\ref{asyG}) we see that light states $\beta M_{\rm phy} \ll 1$ contribute substantially, while contributions from heavy states $\beta M_{\rm phy}\gg 1$ are suppressed by a Boltzmann factor $e^{-\beta M_{\rm phy}}$.

\noindent $\bullet$ 
Taking the low temperature limit of Eq.(\ref{HETthDec4}) we will see that the heterotic string gas at low temperature looks like a particle gas. We set $T=\beta^{-1}\ll 1$, and using the asymptotic behaviors Eq.(\ref{asyG}), we obtain
\begin{align}
	{Z_{\rm het}(\beta) }=\beta V_9\times \beta^{-10}\, \S_0\T_0\, G_{10}(0) +{\cal O}(e^{-\beta})=V_9\,  \S_0\T_0\,c_{10}\times T^{9}+ {\cal O}(e^{-1/T}).
\end{align}
That is, only massless states give non-negligible contribution, while the contribution from massive modes in the mass tower are exponentially suppressed. From the above result we derive the energy density: $\rho= - {1\over V_9} { \partial Z_{\rm het}(\beta) \over \partial \beta}= \S_0\T_0\, 9c_{10}\times T^{10}$, which is the Boltzmann-Steffan law for radiation with ${9\over 2} c_{10}$ the Stefan constant and $2\S_0 \T_0$ the degeneracy. We also have the free energy density $\cf=-{Z_{\rm het}(\beta) \over \beta V_9}= -\, \S_0\T_0\,c_{10}\, T^{10}$, and the pressure $P=-{\partial\, T\,  Z_{\rm het}(\beta) \over \partial V_9}=\S_0\T_0\,c_{10}\, T^{10}$. Also we see that the thermodynamic relation $\rho=T\, \partial_T P-P$ is satisfied.

\noindent $\bullet$ The thermal one-loop amplitude Eq.(\ref{HETthDec4}) is well defined only when the sum over $K$ converges. However we will see that the convergence is ensured only when the temperature is below a critical value, which is just the Hagedorn temperature mentioned in the introduction Sec.\ref{Intro}. Let us see how this happens. Referring to the results in \cite{Bowick:1985az,UniBeh} we have the following asymptotic behavior of the degeneracies as $K \rightarrow \infty$: 
\begin{align} \label{asyS}
	\S_K \! \sim\,  & {\rm const.} \times \left(2\sqrt K \right)^{-11/2} \, \exp\left(2\pi \times {\sqrt 2\over 2} \times 2\sqrt K \right);\\ \ \T_K \! \sim\, & {\rm const.} \times \left(2\sqrt K \right)^{-11/2} \, \exp\left(2\pi \times 1 \times 2\sqrt K\right). \label{asyT}
\end{align}
Since the mass of states are $2\sqrt K$, the above expressions show that the degeneracies grow exponentially with mass. Using also the asymptotic behavior Eq.(\ref{asyG}), we see that in Eq.(\ref{HETthDec4}) the terms in the sum over $K$ are proportional to $K^{-3}\exp\! \left[ \left(2\pi+\sqrt2 \pi -\beta \right) 2\sqrt K\right]$ for large $K$. Then obviously the sign of $\left(2\pi+\sqrt2 \pi -\beta \right)$ in the exponential should be nonpositive in order that the sum over $K$ converge, which amounts to requiring $\beta > \pi \left(2+\sqrt 2 \right)$. This defines the Hagedorn temperature $T_H^{\rm het}=1/\beta_H^{\rm het}$, given by:
\begin{align}
	\beta_H^{\rm het}=\pi \left(2+\sqrt 2 \right). \label{HEThgd}
\end{align}
When the temperature of the gas is above $T_H^{\rm het}$, Eq.(\ref{HETthDec4}) becomes pathologic and the canonical ensemble formalism breaks down. Actually the Hagedorn temperature generically exists in ideal gases of different types of critical strings, whose value does not depend on the dimensionality of spacetime \cite{UniBeh}. The detailed properties of ideal string gases near the Hagedorn temperature and their cosmological consequences will be of interest to investigate, and we will discuss more this issue in Sec.\ref{GDisc}.

\vspace{5mm}



\subsection{Example 2: ideal heterotic string gas in 9 dimensions and its phase structure} \label{9dim}


In this part we consider the effect of compactification on string thermodynamics. In particular, we will show that some phase structure can be induced by compactification, rendering the free energy moduli-dependent.

We illustrate the idea with the simplest model at hand: the compactification of heterotic string on a circle $S^1(R_9)$ in the 9-th direction, with $R_9$ the radius of the circle. Thus the thermal background space is $S^1(R_0)\times \mathbb R^8 \times S^1(R_9)$, where $S^1(R_0)$ is the Euclidean time circle. We can now write down the thermal one-loop amplitude, which is obtained by modifying Eq.(\ref{HETolthm}) according the prescription given in Sec.\ref{ExNonTh} about toroidal compactification. The expression is
\begin{align}
	Z_{\rm het}^{\rm 9d}(\beta) =  {\beta V_8 \over 2(2\pi)^{9}}  \int_{\mfs F} & {d\tau_1\, d\tau_2 \over \tau_2^{11/2}}\, \sum_{a,b} \left[ \sum_{\t m_0,n_0}(-1)^{a\t m_0+bn_0+ \t m_0 n_0}\, e^{-{\pi R_0^2 \over \tau_2}|\t m_0-\tau n_0 |^2} \right]\nnR& \times {1\over \eta^8 \b\eta^8}\, (-1)^{a+b+ab}\,  {\theta[^a_b]^4\over \eta^4} {\b \Gamma_{\rm int} \over \b \eta^{16}}\ \times\! \sum_{m_9,n_9} q^{{1\over 4} \left({m_9\over R_9}-n_9 R_9\right)^2}\b q^{{1\over 4} \left({m_9\over R_9}+n_9 R_9\right)^2}. \label{HETth9D} 
\end{align}
Here the momentum and winding numbers are subscripted with $0$ and $9$ to distinguish the time direction from the 9-th spatial direction.  There is a clash of notation: $V_8$ outside the integral is the volume of the 8 dimensional space, not to be confused with the $SO(8)$-character Eq.(\ref{SO8chV}).

Performing the unfolding trick to Eq.(\ref{HETth9D}) to the sum over $(\t m_0,n_0)$ as what we did to obtain Eq.(\ref{HETthDec}), we have the following decomposition
\begin{align}
	Z_{\rm het}^{\rm 9d}(\beta) = & {\beta V_8 \over 2(2\pi)^{9}}  \int_{\mfs F}  {d\tau_1\, d\tau_2 \over \tau_2^{11/2}}\, \sum_{a,b}  {1\over \eta^8 \b\eta^8}\, (-1)^{a+b+ab}\,  {\theta[^a_b]^4\over \eta^4} {\b \Gamma_{\rm int} \over \b \eta^{16}} \sum_{m_9,n_9} q^{{1\over 4} \left({m_9\over R_9}-n_9 R_9\right)^2}\b q^{{1\over 4} \left({m_9\over R_9}+n_9 R_9\right)^2}\nnR &\hspace{1cm}+ {\beta V_8 \over 2(2\pi)^{9}}  \int_{\sqcup}  {d\tau_1\, d\tau_2 \over \tau_2^{11/2}}\, \sum_{a,b} \left[ \sum_{\t m_0 \neq 0}(-1)^{a\t m_0}\, e^{-{\pi R_0^2 \over \tau_2}\t m_0^2} \right]\nnR& \hspace{2cm} \times {1\over \eta^8 \b\eta^8}\, (-1)^{a+b+ab}\,  {\theta[^a_b]^4\over \eta^4} {\b \Gamma_{\rm int} \over \b \eta^{16}}\! \sum_{m_9,n_9} q^{{1\over 4} \left({m_9\over R_9}-n_9 R_9\right)^2}\b q^{{1\over 4} \left({m_9\over R_9}+n_9 R_9\right)^2}
\end{align}
In terms of $SO(8)$-characters,
\begin{align}
	Z_{\rm het}^{\rm 9d}(\beta)=  & {\beta V_8 \over 2(2\pi)^{9}}  \int_{\mfs F}  {d\tau_1\, d\tau_2 \over \tau_2^{11/2}} {V_8-S_8\over \eta^8 \b\eta^8} {\b \Gamma_{\rm int} \over \b \eta^{16}} \sum_{m_9,n_9} q^{{1\over 4} \left({m_9\over R_9}-n_9 R_9\right)^2}\b q^{{1\over 4} \left({m_9\over R_9}+n_9 R_9\right)^2}\nnR &\hspace{2mm} + {\beta V_8 \over 2(2\pi)^{9}}  \int_{\sqcup}  {d\tau_1\, d\tau_2 \over \tau_2^{11/2}} \left[ \sum_{\t m_0\neq 0} e^{-{\pi R_0^2 \over \tau_2}\t m_0^2} \right] {V_8-(-1)^{\t m_0}S_8 \over \eta^8} {\b \Gamma_{\rm int} \over \b \eta^{24}}\! \sum_{m_9,n_9} q^{{1\over 4} \left({m_9\over R_9}-n_9 R_9\right)^2}\b q^{{1\over 4} \left({m_9\over R_9}+n_9 R_9\right)^2}\nnR
	= &{\beta V_8 \over (2\pi)^{9}}  \int_{\sqcup}  {d\tau_1\, d\tau_2 \over \tau_2^{11/2}} \left[ \sum_{\t \jmath_0} e^{-{\pi R_0^2 \over \tau_2}(2 \t \jmath_0+ 1)^2} \right] {V_8 +S_8 \over 2\eta^8 } {\b \Gamma_{\rm int} \over \b \eta^{24}}\! \sum_{m_9,n_9} q^{{1\over 4} \left({m_9\over R_9}-n_9 R_9\right)^2}\b q^{{1\over 4} \left({m_9\over R_9}+n_9 R_9\right)^2}. \label{Unf9D} 
\end{align}
where in proceeding to the last step, we discarded the $(\t m_0,n_0)=(0,0)$ orbit, as well as the terms with even $\t m_0$ in the $(\t m_0,0)$-orbit, since they are numerically zero ($V_8-S_8=0$). Then we plug in the expansions (\ref{Expan1}) and (\ref{Expan2}), and obtain
\begin{align}
	\!\!\!\!\!\! Z_{\rm het}^{\rm 9d}(\beta) =  &{\beta V_8 \over (2\pi)^{9}}  \int_{\sqcup}  {d\tau_1\, d\tau_2 \over \tau_2^{11/2}} \sum_{\t \jmath_0} e^{-{\pi R_0^2 \over \tau_2}(2 \t \jmath_0+ 1)^2} \!\! \!\!\!\! \sum_{\substack{K\geq 0,\b L\geq -1\\ m_9,n_9\in \mathbb Z}}  \!\! \! \!\!\! \S_K \T_{\b L}\,\, q^{K+{1\over 4} \left({m_9\over R_9}-n_9 R_9\right)^2}\b q^{\, \b L+{1\over 4} \left({m_9\over R_9}+n_9 R_9\right)^2} \nnR 
	 =   &{\beta V_8 \over (2\pi)^{9}}\!\! \int_{\sqcup}  {d\tau_1\, d\tau_2 \over \tau_2^{11/2}}  \sum_{\t \jmath_0} e^{-{\pi R_0^2 \over \tau_2}(2 \t \jmath_0+ 1)^2} \!\! \!\!\!\! \!\!\! \sum_{\substack{K\geq 0,\b L\geq -1\\ m_9,n_9\in \mathbb Z}} \!\! \! \!\!\! \S_K \T_{\b L}\,\, e^{2\pi i \tau_1\left( K-\b L-m_9n_9 \right)}\,\, e^{- \pi \tau_2 \left(2K+2\b L+ {m_9^2\over R_9^2}+n_9^2 R_9^2 \right) }\, ,  \label{Unf9D1}
\end{align}
where in the second line we used the definition $q=e^{2\pi i\tau}$. We notice that in the first exponential the factor $2\pi i\tau_1$ is multiplied by an integer $K-\b L-m_9n_9$. Therefore only terms with $K-\b L-m_9n_9=0$ would survive the integral over $\tau_1$. Thus we have
\begin{align}
	 Z_{\rm het}^{\rm 9d}(\beta) =  &{\beta V_8 \over (2\pi)^{9}}  \int_0^{\infty}\!\!  {d\tau_2 \over \tau_2^{11/2}} \sum_{\t \jmath_0} e^{-{\pi R_0^2 \over \tau_2}(2 \t \jmath_0+ 1)^2} \!\! \!\!\!\!  \sum_{\substack{K\geq 0,\b L\geq -1;\, m_9,n_9\in \mathbb Z\\ K-\b L=m_9n_9}} \!\! \! \!\!\! \S_K \T_{\b L}\,\, e^{- \pi \tau_2 \left(2K+2\b L+ {m_9^2\over R_9^2}+n_9^2 R_9^2 \right) } \nnR =&  \beta V_8\, \times \beta^{-9} \!\! \! \!\!\! \sum_{\substack{K\geq 0,\b L\geq -1;\, m_9,n_9\in \mathbb Z\\ K-\b L=m_9n_9}} \!\! \! \!\!\! \S_K \T_{\b L}\, \, G_9\left( \beta\, \sqrt{2K+2\b L+ {m_9^2\over R_9^2}+n_9^2 R_9^2} \,\right), \label{Unf9D2} 
\end{align}
where $G_d(\,\cdot\,)$ is as in Eq.(\ref{defG}) or (\ref{GlnSusy}). This shows that the physical states have masses-squared $M_{\rm phy}^2=2K+2\b L+ {m_9^2\over R_9^2}+n_9^2 R_9^2\,$, subjected to the level-matching condition $K-\b L=m_9n_9$, and they have $R_9$-dependence. 
To further on analytically, we consider the low temperature approximation $\beta\gg 1$ of Eq.(\ref{Unf9D2}), where only light states give non-negligible contribution. There are four groups of light states:

$\diamond$ states with $K=\b L=m_9=n_9=0$, which are obviously massless for all value of $R_9$;
 
$\diamond$ pure Kaluza-Klein (KK) modes along $S^1(R_9)$ with $K=\b L=n_9=0$ and $m_9\neq 0$, of masses $\left| {m_9\over R_9} \right|$, which become light when $R_9\rightarrow \infty$;

$\diamond$ pure winding modes along $S^1(R_9)$ with $K=\b L=m_9=0$ and $n_9\neq 0$, of masses $\left| n_9 R_9 \right|$, which become light as $R_9\rightarrow 0$;

$\diamond$ states with one unit of momentum and winding along $S^1(R_9)$: $m_9=n_9=\pm 1$, with $K=0$, $\b L=-1$, whose masses are $\left|R_9^{-1}-R_9\right|$, becoming massless when $R_9=1$, responsible for the gauge symmetry enhancement $U(1)\rightarrow SU(2)$.

All other physical states have masses of at least one unit of $l_s^{-1}$ for all value of $R_9$, whose contribution to Eq.(\ref{Unf9D2}) are suppressed at least by $e^{-\beta}$. Therefore in the summation in Eq.(\ref{Unf9D2}) we only need to conserve the terms associated to the light states enlisted above, which yields
\begin{align}
	{ Z_{\rm het}^{\rm 9d}(\beta) } =  & \,\beta V_8\times \beta^{-9}\, \bigg\{\S_0\T_0 \,G_9(0)+ \S_0\T_0 \sum_{m_9\neq 0}G_9\! \left(\beta\, |m_9| /R_9 \right) \nnR & \hspace{5mm} + \S_0\T_0 \sum_{n_9\neq 0} G_9 \! \left(\beta\, |n_9|R_9\right)
	 + 2\, \S_0\T_{-1} \, G_9 \!\left(\beta\, \left| R_9^{-1} -R_9\right|\right) \bigg\}+ {\cal O}(e^{-\beta}). \label{Unf9D4} 
\end{align}
To show the phase structure, we study the density of the Helmholtz free energy $\cf_{\rm het}^{9d}=-Z_{\rm het}^{9d} (\beta)/\beta V_8$. Recalling that $G(|x|)$ is significantly nonzero in the neibourhood of $x=0$ and that it decreases exponentially as $|x|$ increases, therefore the free energy density are well approximated by the following expressions in different regions of $R_9$:
\begin{align}
	\cf_{\rm het}^{9d}\approx \left\{\begin{array}{lll}
		\displaystyle -T^9 \Big[\S_0\T_0 \,G_9(0)+2\, \S_0\T_{-1} \, G_9 \!\left(\beta\, \left| R_9^{-1} -R_9\right|\right)\Big], & |R_9^{-1}-R_9|<R_0^{-1}\,, &\text{phase I}; \\ \vspace{-3mm} \\
		\displaystyle -T^9 \Big[ \S_0\T_0 \,G_9(0)+\S_0\T_0 \sum_{m_9\neq 0}G_9\! \left(\beta\, |m_9| /R_9 \right) \Big], & R_9>R_0\, , &\text{phase III};  \\  \vspace{-3mm} \\
		\displaystyle -T^9 \Big[ \S_0\T_0 \,G_9(0)+\S_0\T_0 \sum_{n_9\neq 0}G_9\! \left(\beta\, |n_9| R_9 \right) \Big], & R_9<R_0^{-1} \,, &\text{phase III'}; \\  \vspace{-3mm} \\
		\displaystyle -T^9\, \S_0\T_0 \,G_9(0), & {\rm otherwise}, &\text{phase II and II'}.
	\end{array}\right.  \label{FreeEn} 
\end{align}
Plotting this free energy density against $\ln \!R_9$ at fixed temperature we obtain Fig.\ref{ptHet}, where the reason to choose $\ln\! R_9$ as variable is to display the symmetry under T-duality $R_9\rightarrow 1/R_9$. We can distinguish different phases where the string gas behaves differently, numbered with Roman numerals both in Fig.\ref{ptHet} and in Eq.(\ref{FreeEn}):

\begin{figure}[t]
\hspace{0mm}\scalebox{0.92}{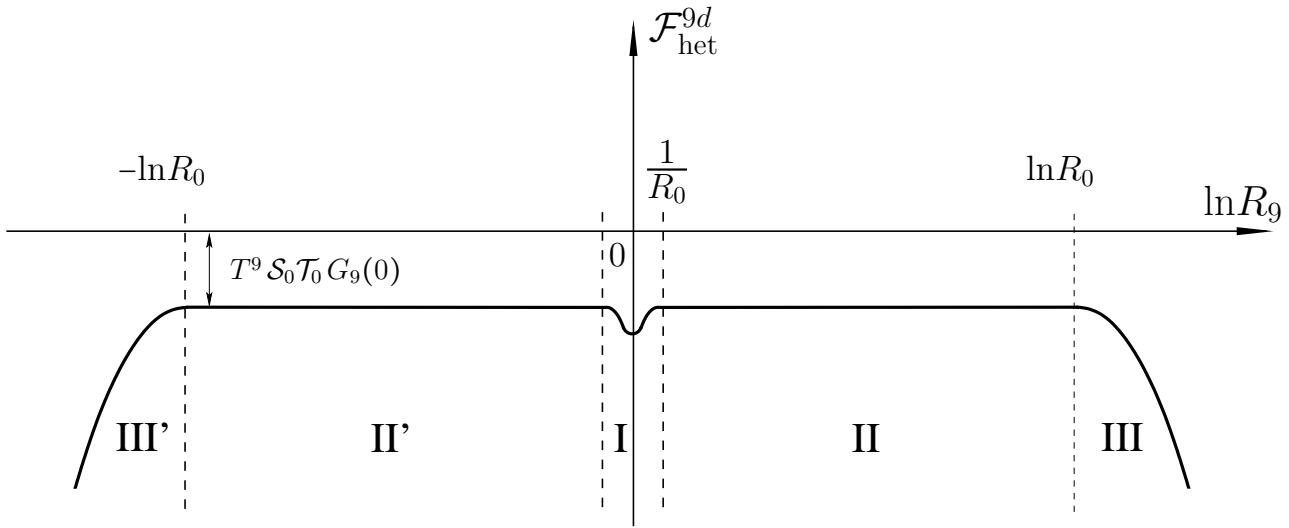}
\caption{\small Schematic plot of the low temperature free energy density Eq.(\ref{FreeEn}) of an ideal heterotic string gas compactified on the circle $S^1(R_9)$.  We use $\ln \! R_9$ in order that the T-duality symmetry is manifest.} \label{ptHet}
\end{figure}

$\diamond$ In the phase I we see a local minimum at $R_9=1$, to which $R_9$ can be attracted, and this provides the mechanism for stabilizing the modulus $R_9$.  

$\diamond$ In the phases III and III' the radius $R_9$ is attracted to $\infty$ and $0$ respectively, so that the system is decompactified.

$\diamond$ In the phases II and II' $R_9$ can be fixed at any value. 

\noindent The following observations can be drawn from the computations we have gone through so far:

\noindent $\bullet$ The free energy density is lowered whenever there are extra light or massless states appearing in the spectrum. In phase I, we have the local minimum induced by the $4\S_0\T_{-1}$ light states of mass $|R_9^{-1}-R_9|$, which carry one unit of momentum and one unit of winding along $S^1(R_9)$; in phase III the slope is induced by the pure KK modes of masses $|m_9|/R_9$ ($m_9\neq 0$); in phase III' the slope is induced by the pure winding modes of masses $|n_9|R_9$ ($n_9\neq 0$). Generically for the models mentioned in Sec.\ref{ExNonTh}, the Helmholtz free energy density at one-loop is always lowered when extra massless or light states emerge in the spectrum, and thus at weak coupling, thermal effects attract the system to nearby available configurations with more massless or light states.

\noindent $\bullet$ Whereas in the previous example in Sec.\ref{10dim} without compact spatial dimension, stringy effects were manifest only at high temperature (where we have observed the Hagedorn temperature), here upon compactification, strictly stringy effects can show up at low temperature, because we can have light winding modes which are thermalized at low temperature. The stringy effects observed in Fig.\ref{ptHet} are: the local minimum in phase I induced by states with both winding and momentum numbers; the slope in phase III' induced by pure winding modes. Also the winding modes are essential to implementing the symmetry under T-duality. Similar features related to the winding modes can be observed in other types of strings mentioned in Sec.\ref{ExNonTh}. 

In Sec.\ref{COSapp} we will investigate the cosmological implications of this phase structure. More complex phase structure can be produced when more sophisticated compactifications are implemented, producing a larger moduli space.

\section{General discussion about high temperature behavior} \label{GDisc}




In this section we pick up the topic of Hagedorn temperature suspended by the end of Sec.\ref{10dim}. The discussions to be presented apply generically to the class of ideal string gases that we consider: those of critical strings living in flat spacetime with toroidal or orbifold compactifications.

We will first show that the presence of a Hagedorn temperature is general in the string gases of our concern. Then we will show with explicit calculation, that some thermal winding modes can become tachyonic above the Hagedorn temperature, and they are responsible for the breaking down of the canonical ensemble description. We will hence be led to address the aspect that the Hagedorn temperature can be a phase transition point. In the last part of this section we will discuss the conditions for the onset of the Hagedorn phase transition, the Hagedorn instability viewed from microcanonical ensemble, and also the current understanding of the nature of this phase transition.

\subsection{Hagedorn temperature in different types of ideal string gases} \label{Sec41}


Here we explain how the Hagedorn temperature emerges in generic ideal string gases. The thermal one-loop amplitude has the following general form,
\begin{align}
	Z(\beta) :=\ln \!\left( {\rm Tr}\, e^{-\beta H} \right)=Z_{0}(\beta)+Z_1(\beta), \label{Z0Z1}
\end{align}
where $Z_0$ is the non-thermal vacuum bubble contribution, i.e. the Casimir energy, and $Z_1$ is the piece accounting for thermal effect which vanishes when $\beta\rightarrow \infty$. The examples that we studied fit into Eq.(\ref{Z0Z1}): for instance  Eq.(\ref{HETthDec}) and the first two lines of Eq.(\ref{Unf9D}), where the first line is $Z_0$ and the second line $Z_1$. The Casimir energy $Z_0(\beta)$ is just the non-thermal one-loop amplitude addressed in Sec.\ref{NonThAmpSec}, which is free of UV divergence and is irrelevant to the Hagedorn temperature. Then we examine the thermal piece $Z_1(\beta)$, which can be written as
\begin{align}
	Z_1(\beta)={ V_{d-1} \over (2\pi)^{d-1}} \int_0^{\infty}\!\! dM \! \! \! \int \!\! \! d\vec p\, \bigg[ -\rho_{\rm b}(M) \ln \! \left(\! 1-e^{\, -\beta \sqrt{\vec p^2+M^2}}\right)+\rho_{\rm f}(M)\ln\!\left(\! 1+e^{\, -\beta \sqrt{\vec p^2+M^2}} \right) \bigg]\, , \label{Z1Z1}
\end{align}
where $\vec p$ is the $(d-1)$-dimensional spatial momentum, $M$ runs over the mass spectrum of one single string, and $\rho_{\rm b/f}(M)$ is the density of states of one bosonic/fermionic string at mass $M$. In case of unbroken spacetime supersymmetry such as the examples in Sec.\ref{10dim} and Sec.\ref{9dim}, we have $\rho_{\rm b}(M)=\rho_{\rm f}(M)$. Then by Eq.(\ref{GlnSusy}), we recover the expressions with the function $G_d(\,\cdot\,)$ as in Eqs (\ref{HETthDec4}) and (\ref{Unf9D2}).

%
%
%

According to the results in \cite{UniBeh}, the densities of states have the following asymptotic behavior for large mass $M$ for a closed string in $d$ dimensions at weak coupling:
\begin{align}
	\rho_{\rm b,f} (M) \sim {\rm const.}\times M^{-d} \, e^{\, 2\pi \left( \sqrt {\omega_{\rm L}}+ \sqrt {\omega_{\rm R}} \right) M}, \ \ \ \ (\text{for closed strings}), \label{asyD}
\end{align}
where $(\omega_{\rm L},\omega_{\rm R})=(1,1)$, $(\demi, \demi)$ and $(\demi, 1)$ for bosonic, type II and heterotic strings respectively. On the other hand using Eq.(\ref{FMLA8}) and the asymptotic behavior of the modified Bessel function of the second kind, we have
\begin{align}
	\pm \int\!\! d\vec p\, \ln \left(1\pm e^{-\beta \sqrt{\vec p^2 + M^2}}\right)\sim  {\rm const.}\times (T M)^{d-1 \over 2} e^{-\beta M}, \ \ {\rm when}\ \ \beta M \gg 1. \label{asyGpm}
\end{align}
Plugging this and Eq.(\ref{asyD}) in Eq.(\ref{Z1Z1}), we find that the integration over $M$ becomes divergent when $\beta$ is below the critical value $2\pi\, \left(\!\sqrt{ \omega_{\rm L}}+ \sqrt{ \omega_{\rm R} }\right)$, which is just the inverse Hagedorn temperature $\beta_H$. The explicit values for different types of string theories are
\begin{align}
	\beta_H:=2\pi \left(\sqrt {\omega_{\rm L}} + \sqrt {\omega_{\rm R}} \, \right)=\left\{ \begin{array}{ll}  4\pi & \text{bosonic string}; \\ 2\sqrt 2\, \pi & \text{type II string}; \\ (2+\sqrt 2)\, \pi & \text{heterotic string}. \end{array} \right. \label{HagBIIH}
\end{align}

In the same way we can also investigate models containing open strings. The type I string has the same Hagedorn temperature as the type II string, because {\em i)} the closed string sector is the type II string with states of negative worldsheet parity removed, where the truncation of spectrum does not alter the behavior Eqs (\ref{Z1Z1}) and (\ref{asyD}); {\em ii)} in the open string sector Eq.(\ref{Z1Z1}) applies as well, and the open string counterpart of Eq.(\ref{asyD}) is \cite{UniBeh} 
\begin{align}
	\rho_{\rm b,f}(M)\sim {\rm const.}\times M^{1-d\over 2} e^{\, \beta_H M}, 
	\ \ {\rm when} \ M\rightarrow \infty  \ \ (\text{for open strings}), \label{asyDOp}
\end{align}
where the inverse Hagedorn temperature $\beta_H$ is the same as in the closed string sector, $\beta_H=2\sqrt 2 \,\pi$. The same logic applies to the open bosonic string where we have Eq.(\ref{asyDOp}) with Hagedorn temperature given by $\beta_H=4\pi$, identical to that of the closed bosonic string. 

We have two further points to make: 

\noindent $\bullet$ The Hagedorn temperature does not depend on compactification, but only on the type of string theory. This can be seen from Eqs (\ref{asyD}) and (\ref{asyDOp}), where the coefficient of $M$ in the exponential, which defines the inverse Hagedorn temperature, does not depend on the spacetime dimension $d$ but only on the type of the string.

\noindent $\bullet$ The behavior of an ideal string gas near the Hagedorn temperature can depend not only on the type of the string, but also on the spacetime dimension. The power of $M$ dressing the exponential in Eqs (\ref{asyD}) and (\ref{asyDOp}) depend on $d$, and different $d$ will result in different convergence properties of the integration over $M$ in Eq.(\ref{Z1Z1}) at the point $\beta=\beta_H$. More details will be covered in Sec.\ref{Sec43}.

\subsection{Thermal tachyons above the Hagedorn temperature} \label{Sec42}

For closed string gases in $d$ dimensions, investigating their Hagedorn temperature from the point of view of the $(d-1)$-dimensional effective theory can lead to another level of insight, where the reduced dimension is the Euclidean time. Basically, the $(d-1)$-dimensional effective theory contains states whose masses depend on temperature, where some winding states along the Euclidean time circle can become tachyonic when the temperature exceeds the Hagedorn temperature. This causes the one-loop vacuum-to-vacuum amplitude to diverge literally, implying that near the Hagedorn temperature some instability can set in. We will discuss this aspect in the next subsection while for the moment we explain how the thermal tachyons emerge.


On the technical level, the starting point is still the thermal one-loop amplitude. However now considering the thermal gas in a $(d-1)$-dimensional effective theory, we need to characterize the compact Euclidean time $S^1(R_0)$ with a lattice sum, something looking like Eq.(\ref{GmTo}), instead of an instanton sum as Eq.(\ref{instII}) or (\ref{instHET}) which has been the case in all the previous calculations. Noticing that the the Euclidean time is not compactified toroidally but \`a la Scherk-Schwarz, a freely-acting orbifold compactification, therefore the resulting lattice sum is more complicated than that from a toriodal compactification. It turns out that the lattice is shifted differently following different combinations of spin structures, that is, the internal momenta $\{p_{\rm L,R}^I\}$ in Eq.(\ref{GmTo}) become functions of spin structure indices: the $(a,b)$ and $(\b a,\b b)$ as in Eqs (\ref{IIABol}) and (\ref{HETol}) for instance. The detailed construction of such lattice sums can be found many literatures (c.f. for example \cite{Kounnas:1989dk, Antoniadis:1991kh, Antoniadis:1999gz, Angelantonj:2008fz}), while a shortcut available for us here is to by performing a Poisson resummation to the instanton sum. 

To be more explicit, we invoke the simplest model that we have studied, the heterotic string in 10 dimensions, whose thermal one-loop amplitude in terms of worldsheet instaton sum is already obtained in Eq.(\ref{HETolthm}). We proceed by Poisson resumming the index $\t m$ using Eq.(\ref{form9}), where we need to put $(-1)^{a\t m}=e^{\, i\, \pi\, a\t m}$ and combine this exponential with $e^{-{\pi R_0^2 \over \tau_2}|\t m-\tau n|^2}$. The computation is slightly lengthy but straightforward, and the result is effectively a non-thermal one-loop amplitude in 9 dimensions:
\begin{align}
	Z_{\rm het}(\beta) =  { V_9 \over 2(2\pi)^9 }\! \int_{\mfs F} {d\tau_1 d\tau_2 \over \tau_2^{11/2}} \, \bigg\{ {O_8\over \eta^8 }{\Gamma_{\rm int}\over \b \eta^{24}}\, \sum_{m,n}q^{{1\over 4}\left[ {m+\demi\over R_0}-(2n+1) R_0\right]^2} \b q^{{1\over 4}\left[ {m+\demi\over R_0}+(2n+1) R_0\right]^2}+\dots \bigg\} . \label{HetTLat}
\end{align}
Here we have only written down the term relevant to the Hagedorn temperature which can develop divergence for certain values of $R_0$. The omitted terms give rise to analytic results integrated over $\mfs F$ for all real value of $R_0$. Also we notice that the lattice involved above is shifted with respect to the toroidal case given by Eq.(\ref{GmTo}) with Eq.(\ref{p9LR}), in that $m\rightarrow m+\demi$ and $n\rightarrow 2n+1$. Using the power expansion (\ref{SO8chO}) together with the definition of the $\eta$-function Eq.(\ref{DefEta}), we have 
\begin{align}
	{O_8\over \eta^8 }=q^{-1/2}+36\, q^{1/2}+402\, q^{3/2}+{\cal O}(q^{5/2}). \label{OEtaExp}
\end{align}
Using also Eq.(\ref{Expan2}), we have the following expansion of the integrand in Eq.(\ref{HetTLat})
\begin{align}
	{O_8\over \eta^8 }{\Gamma_{\rm int}\over \b \eta^{24}}\, \sum_{m,n}q^{{1\over 4}\left[ {m+\demi\over R_0}-(2n+1) R_0\right]^2} \b q^{{1\over 4}\left[ {m+\demi\over R_0}+(2n+1) R_0\right]^2}= 2\, (q\b q)^{{1\over 4} \left(R_0^2+{1\over 4 R_0^2}-3\right)} + ({\rm higher\ orders}), \label{QQMst}
\end{align}
where again we exhibited only the term responsible for the definition of Hagedorn temperature, since all other terms are convergent integrated over $\mfs F$ for all real value of $R_0$. This term comes from the leading terms in Eqs (\ref{Expan2}) and (\ref{OEtaExp}), and the two terms in the lattice sum of $(m,n)=(0,0)$ and $(-1,-1)$. Therefore it is contributed by the modes corresponding to closed string winding once and carrying half unit of momentum along the Euclidean time circle, where coefficient 2 accounting for the fact that there are two of them: $(m,n)=(0,0)$ and $(-1,-1)$. These thermal winding modes have temperature-dependent mass that we denote by $M_*$, given by the power of $q\b q$ in Eq.(\ref{QQMst}):
\begin{align}
	M_*^2=R_0^2+{1\over 4 R_0^2}-3={1\over (2\pi T)^2}+(\pi T)^2-3. \label{TWms}
\end{align}
Displaying in Eq.(\ref{HetTLat}) only the terms corresponding to these thermal winding states, we have
\begin{align}
	Z_{\rm het}(\beta) =  { V_9 \over 2(2\pi)^9 }\! \int_{\!\mfs F} \! {d\tau_1 d\tau_2 \over \tau_2^{11/2}} \, 2\, (q\b q)^{{1\over 4} \left(R_0^2+{1\over 4 R_0^2}-3\right)} +\, \dots\, = { V_9 \over 2 (2\pi)^9 }\! \int_{\! \mfs F} \!{d\tau_1 d\tau_2 \over \tau_2^{11/2}} \,  2\,e^{- \pi\tau_2 M_*^2}  + \dots \label{HetTLat2}
\end{align}
With Eq.(\ref{TWms}) we notice that when $1-1/\sqrt 2<R_0<1+1/\sqrt 2$ the thermal winding states of our concern are tachyonic ($M_*^2<0$), rendering the integral over $\mfs F$ above divergent so that the canonical ensemble description breaks down. Therefore if we start from very low temperature (large $\beta$), and heat the gas gradually ($\beta$ reduces), we will find the critical point where the canonical ensemble breaks down is $\beta=\pi (2+\sqrt 2)$. Unsurprisingly this critical value is exactly the inverse Hagedorn temperature $\beta_H^{\rm het}$ obtained from 10-dimensional point of view Eq.(\ref{HEThgd}). 

Generically when we carry out the same analysis to other types of closed strings in $d$-dimensions that we have discussed in Sec.\ref{NTHtoTH}, the resulting one-loop amplitudes in the $(d-1)$-dimensional effective theory contain systematically a term of the form
\begin{align}
	({\rm degeneracy})\times {V_{d-1}\over 2(2\pi)^{d-1}}\int_{\mfs F} {d\tau_1 d\tau_2 \over \tau_2^{d+1 \over 2}}\, e^{-\pi \tau_2\, M_*(\beta)^2 }, \label{dm1}
\end{align}
induced by some thermal winding modes, whose mass-squared $M_*(\beta)^2$ is positive for large $\beta$, but becomes negative as $\beta$ decreases (or the temperature $T$ increases) till below a critical value. This critical value is just the inverse Hagedorn temperature $\beta_H$, which matches the results in Eq.(\ref{HagBIIH}).

\subsection{Hagedorn instability and phase transition} \label{Sec43}

In this part we come back to the point about the Hagedorn temperature as an embarkation point for the ideal string gas to a new phase. In the following we will first examine the behavior of ideal string gases approaching the Hagedorn temperature, which is important for judging whether the phase transition can happen. Then we will comment on the microcanonical approach to the Hagedorn temperature and its link with canonical description. Finally we will give some qualitative discussions about the phase transition itself.





\subsubsection*{Limiting and non-limiting Hagedorn temperature}



The Hagedorn temperature of an ideal string gas can be qualified as {\em non-limiting} or {\em limiting} according to whether or not the string gas can reach the Hagedorn temperature by absorbing a finite amount of energy \cite{AlvarezStFT,Dienes:1998hx}. The Hagedorn phase transition cannot happen for limiting cases since the phase transition point cannot be reached in practice.

Restricted to the models of our interest, we can verify, using Eqs (\ref{Z1Z1}) to (\ref{asyGpm}), that the thermal one-loop amplitude $Z(\beta)=\ln {\rm Tr}\, e^{-\beta H}$ at the limit $\beta\rightarrow \beta_H^+$: 

\noindent $\diamond$ tends to a finite value for closed string gas in $d\geq 2$ dimensions;

\noindent $\diamond$ tends to infinity for closed string gas in $d=1$ dimension and open string gas in all dimensions,

\noindent which are inferred from the convergence property of the integral over $M$ in Eq.(\ref{Z1Z1}). For closed string gases, this can also be verified from the viewpoint of the $(d-1)$-dimensional effective field theory: at $\beta=\beta_H$ ($M_*(\beta_H)^2=0$) the infinite integration over $\tau_2$ in Eq.(\ref{dm1}) converges for $d\geq 2$ and diverges for $d=1$. We also notify that the presence of D-branes in the background can affect the limiting behavior of a string gas at $\beta\rightarrow \beta_H^+$, which can be studied quantitatively (c.f. for example \cite{Abel:1999rq} and the references therein). 

Thus we can easily derive the behavior of the total energy $E(\beta)=-{\partial Z(\beta) \over \partial \beta}$. Taking into account the impact of the derivative with respect to $\beta$, we find

\noindent $\diamond$ $E(\beta)$ tends to a finite value when $\beta_H$ for ideal closed string gases in $d\geq 4$ $\Rightarrow$ the Hagedorn temperature is non-limiting, which is a site of phase transition; 

\noindent $\diamond$ $E(\beta)$ diverges when $\beta\rightarrow \beta_H^+$ for ideal closed string gases in $d\leq 3$ and ideal open string gas $\Rightarrow$ the Hagedorn temperature is limiting, which is the maximum temperature of the gas.

\noindent In the limiting case, one can never achieve Hagedorn temperature attempting to heat an ideal string gas by pumping energy into it. In fact as the more and more energy is concentrated in the string gas, some instability other than the Hagedorn instability would already set in and turn the homogeneous string gas into something else, for example a black hole.

\subsubsection*{Microcanonical ensemble approach}

We have seen in the previous discussions, that for the non-limiting cases, the string gases can go beyond the Hagedorn temperature at a finite cost of energy, but the canonical ensemble fails to trace the string gases across the Hagedorn temperature from below since it becomes ill defined. Therefore it is of interest to try the more fundamental microcanonical ensemble description, which can hopefully uncover properties of string gases beyond the Hagedorn temperature. 

To this end we need to compute $\Omega(E)$, the number of microstates of the string gas. For the non-limiting cases (closed strings in $d\geq 4$ dimensions), it is possible to obtain the asymptotic behavior of $\Omega(E)$ at large energy density using the technique in \cite{HagInst} for bootstrap models. We can find that for closed strings with densities of states given by Eq.(\ref{asyD}) and with $d\geq 4$, the bootstrap condition is satisfied: $\rho_{\rm b,f}(E)\sim \Omega(E)$ for large enough energy density \cite{Sundborg:1984uk,Bowick:1985az}. Thus we have for the non-limiting cases $\Omega(E)\sim E^{-d}\, e^{\beta_H E}$ ($d\geq4$, $E\rightarrow \infty$). This allows the computation of the temperature $T=\left({\partial \ln \Omega(E) \over \partial E}\right)_V^{-1}$ and the specific heat $C_V=\left({\partial E\over \partial T}\right)_V$, which yields a temperature higher than the Hagedorn temperature with the specific heat negative. This shows in the microcanonical language that the system is unstable when its temperature exceeds the Hagedorn temperature, and this instability should reasonably be identified the Hagedorn instability. 

Another way of computing $\Omega(E)$ is to start from the canonical partition function ${\rm Tr}\, e^{-\beta H}=\exp\! \left[Z(\beta)\right]$ and reverse the relation ${\rm Tr}\, e^{-\beta H}= \int\! dE\,  \Omega(E)\, e^{-\beta E}$ which is just a Laplace transform. Actually the thermal one-loop amplitude $Z(\beta)$ can be mathematically defined beyond the Hagedorn temperature  $\beta<\beta_H$ for closed strings, although literally the integral over $M$ in Eq.(\ref{Z1Z1}) evaluates to infinity. This can be seen explicitly in Eq.(\ref{dm1}), where the integration can be defined for $M_*(\beta)^2<0$ (where $\beta<\beta_H$) through analytical continuation. In fact, $Z(\beta)$ as a function of $\beta$ can be analytically continued to the whole complex plane where $\beta_H$ is the rightmost branch point at which $Z(\beta)$ can either be regular or singular \cite{Deo:1988jj}. The asymptotic behavior of $Z(\beta)$ approaching $\beta_H$ allows us to compute the asymptotic behavior of $\Omega(E)$ at large energy density, and the result for string gases with non-limiting Hagedorn temperature is just as in the last paragraph. More details of this approach can be found in \cite{Deo:1988jj,Deo:1989bv,Deo:1991mp}.


We still need to explain one puzzle concerning the relation between the canonical and the microcanonical description: apparently above the Hagedorn temperature the former breaks down while the latter can still be valid. Indeed the equivalence between the canonical and the microcanonical ensemble no longer holds above the Hagedorn temperature, mathematically since the saddle point approximation, the key step proving this equivalence using ${\rm Tr}\, e^{-\beta H}= \int\! dE\,  \Omega(E)\, e^{-\beta E}$, is no longer available for $\beta<\beta_H$, due to the factor $e^{\beta_H E}$ in $\Omega(E)$. One can also compute in the canonical ensemble description the energy fluctuation $\langle (E-\langle E \rangle)^2\rangle \over\langle E \rangle^2$ which shows that it becomes large with the growing of energy density \cite{Bowick:1985az}. Therefore at some point of high energy density and hence of high temperature, the energy fluctuation becomes so large that the canonical ensemble is no longer suitable for describing an isolated string gas and one need to switch to the microcanonical ensemble.

\subsubsection*{Phase transition at Hagedorn temperature}

In the final part of this section we address the issue of Hagedorn phase. The Hagedorn temperature was first unraveled in the dual resonance model of hadrons \cite{Hagedorn:1965st}, where the hadron density $\rho(m)$ at mass $m$ behaves asymptotically as 
\begin{align}
	\rho(m) \sim  c\, m^{-a} e^{\, b\, m}\ \ (m\rightarrow \infty), \label{asyHadron}
\end{align}
where $a,\, b,\, c$ are positive constants. We notice at once that the asymptotic behaviors of single string state densities Eqs (\ref{asyD}) and (\ref{asyDOp}) have the same form, so that in the same way as in string theory, Eq.(\ref{asyHadron}) predicts a critical temperature $T=1/b$ (the genuine Hagedorn temperature) above which the canonical ensemble formalism breaks down. It was then observed that when $a> {5\over 2}$ (in spacetime dimension 4) the hadronic system exhibits instability around the Hagedorn temperature \cite{HagInst}. Later as QCD proved to be the more fundamental theory of strong interaction, it was realized that this instability is the point of phase transition of the hadron system bridging the hadronic phase at low energy and the quark-gluon phase at high energy \cite{Cabibbo:1975ig}.

The work on the Hagedorn instability in hadron systems prompt reflection on the fundamental degrees of freedom of string theory. 
It may turn out that behind the modes in the mass tower, obtained from quantizing the Polyakov worldsheet action, there are more fundamental constituents. The latter are the true degrees of freedom of strings, which do not manifest themselves until the temperature is high enough, just like the quarks and gluons which are liberated at high temperature.


Some interesting speculations have been made in \cite{Atick:1988si}. Based on the thermal effective action, it is argued that the system should undergo a first-order phase transition somewhat below the Hagedorn temperature; in the high temperature phase the free energy $F$ scales as $F \sim T^2$ as if a $(1+1)$-dimensional quantum field theory, implying that string theories in UV contain far fewer degrees of freedom per unit volume than even quantum field theories.  A more detailed analysis is given in \cite{Antoniadis:1991kh}, where the exact thermal effective action is worked out using supersymmetry constraints. However the result shows that the phase transition is not of first order, and that the new phase is a non-critical string theory in ($5+1$) dimensions. Later on string theory dualities were exploited to examine the role of non-perturbative effects in the Hagedorn phase transition \cite{Antoniadis:1999gz}.

Although the Hagedorn instability of ideal string gases received immediate attention in the very beginning stage of string thermodynamics, results obtained so far are based on effective field theories, and it is still difficult to achieve a description at full string level. The issue of Hagedorn phase transition, and its implementation in early cosmology, remains an open question.

\section{Ideal string gases in early cosmology} \label{COSapp}





In this section we address the cosmological implications of the string thermodynamics in the previous sections, based on the work in \cite{cosmoB,cosmoC,stabmod1,stabmod2,cosmoreview,Estes:2011iw,Liu:2011nw,Florakis:2010is,Kounnas:2011fk,Kounnas:2011gz}. In particular we will show that a string scenario for cosmology can be built up from first principles, where the string thermodynamics covered in Sec.\ref{TISG} can be naturally incorporated. We will then discuss the important phenomenological issues and some open questions, where most of the discussions will be qualitative.


%


\subsection{The thermal string scenario and its effective theory}

Due to the nature of string theory as a candidate ``theory of everything'' which offers the matter, the interactions and the spacetime in a whole package, string cosmology is not about how the Einstein universe would evolve once filled up with strings instead of particles. Rather, it is about how cosmological evolution can eventually arise as a solution of string theory among others.

On the practical level, a good point to start with is the the effective field theory, from which we can derive the equations of motion governing the cosmological evolution. In order that the model be analytically tractible, we restrict our attention to {\em critical superstring theory} and we assume {\em weak coupling}. We can therefore adopt a perturbative approach, searching for cosmological solutions order by order. Since we are meant to take into account thermal effects, {\em the effective action should be computed against a thermal background with compact Euclidean time, and then continued back to Lorentzian signature} when we derive the equations of motion.

The effective supergravity at tree level cannot feel the thermal effects since it is computed on a genus-0 Riemann surface which cannot wrap the Euclidean time circle. As a result the tree level effective theory does not lead to cosmological evolution, since generically the effective supergravities give AdS or Minkowski space as vacuum solutions. Therefore higher order corrections need to be investigated\footnote{Certainly, non-perturbative corrections should equally be considered if they contribute substantially.}.

By phenomenological consideration, it is relevant to choose models with Minkowski space as the tree level vacuum since the observed cosmological constant is tiny and positive (roughly $10^{-120}M_{\rm Planck}^4$). Therefore more accurately, the full tree-level background space is
\begin{align}
	\left(\text{Tree-level background space} \right)=\mathbb R^{1,d-1}\times \cm_{\rm int}\, , \label{BG1}
\end{align}
where the visible spacetime is of dimension $d$ and the rest $10-d$ dimensions are compactified on some internal space $\cm_{\rm int}$. When computing higher order corrections at finite temperature, we need to use the corresponding thermal background
\begin{align}
	\left(\text{Thermal tree-level background space} \right)=S^1(\h R_0)\times T^{d-1}(\h R_{\rm box}) \times \cm_{\rm int}\, ,  \label{BG2}
\end{align}
where $S^1(\h R_0)$ is the Euclidean time circle and $T^{d-1}(\h R_{\rm box})$, the spatial part of visible spacetime, is a $(d-1)$-dimensional periodical box with the length of each dimension $2\pi \h R_{\rm box}$, introduced to regularize the space volume: $\h V_{d-1}=(2\pi \h R_{\rm box})^{d-1}$. Here we put hat on $R_0$, $R_{\rm box}$ and $V_{d-1}$ to indicate that they are measured in string frame. Since we will work both in string frame and in Einstein frame, the distinction of the two is necessary.  Therefore from now on, we use the convention that {\em hatted quantities are string frame quantities, and non-hatted quantities are either Einstein frame quantities or quantities indifferent of reference frame}.\footnote{For a quantity, say $L$, of dimension of length, its string frame value $\h L$ and Einstein frame value $L$ are related by a dilaton-dependent rescaling $\h L = e^{{2\over d-2}\phi_d} \, L$ with $d$ the spacetime dimension and $\phi_d$ the dilaton in $d$ dimensions.} Since all the calculations of one-loop amplitudes in the previous two sections were carried out in the string frame, when we use those results here, we need to add hat wherever necessary.

The metric of background (\ref{BG2}) is
\begin{align}
	d\h s^2_{\rm thermal}=\h \beta^2 du^2+\sum_{i=1}^{d-1}(2\pi \h R_{\rm box})^2 (dx^i)^2+ ds_{{}_{}^\cm}^2,\ \ ( 0\leq u, x^i <1). \label{THMetric}
\end{align}
Here $ds_{{}_{}^\cm}^2$ is the metric of the internal space; $u$ and $x^i$ ($i=1,\dots,d-1$) are the Euclidean time and spatial coordinates chosen such that their periods are $1$; $\h \beta=2\pi \h R_0$ the perimeter of the Euclidean time circle is the inverse temperature in the string frame. This choice of tree-level vacuum brings along a bonus: the full string spectrum can be known exactly (since we consider here only toroidal or orbifold $\cm_{\rm int}$ which allow exact solution of the worldsheet CFT), and hence the perturbative quantum corrections at one-loop can be precisely computed. Different from the tree level action, the one-loop correction can feel the temperature since the genus-1 worldsheet (torus for closed strings) has non-contractable loops which can wrap the compact Euclidean time. In fact the one-loop correction computed against the background (\ref{BG2}) is nothing but the thermal vacuum-to-vacuum amplitude studied in Sec.\ref{TISG}.

At this point we can write down the effective action computed to one-loop at finite temperature:
\begin{align}
	S=\int d^dx \sqrt{- \h g}\, \left[e^{-2\phi_d} \left({\h \R \over 2}+2(\partial \phi_d)^2+\dots\right)+ {Z(\h \beta) \over \h \beta \h V_{d-1}}\right], \label{EffAct1L}
\end{align}
which is already continued to Lorentzian signature. The first term in the bracket with dilaton dressing is the tree-level part, where $\h \R$ is the Ricci scalar of the string frame metric $\h g$, $\phi_d$ is the dilaton in $d$-dimensions which is set to be small $e^{\phi_d}\ll1$ to ensure weak coupling, and the ellipses stand for the kinetic terms of other scalar fields, whose content is model-dependent. The second term in the bracket in Eq.(\ref{EffAct1L}) is the one-loop correction, where $Z(\h \beta)$ is just the thermal one-loop amplitude computed in Sec.\ref{TISG}, and it can depend on scalar fields, for example Eq.(\ref{Unf9D2}) has $R_9$-dependence. 

To simplify further discussions, we switch to the Einstein frame by rescaling the metric $\h g=e^{{4\over d-2}\phi_d} g$. Also we notice that ${Z(\h \beta) \over \h \beta \h V_{d-1}}=\h \cf$ is the string frame density of Helmholtz free energy, related to the Einstein frame counterpart by $\cf={Z(\h \beta) \over  \beta  V_{d-1}}= e^{{2d \over d-2}\phi_d}{Z(\h \beta) \over \h  \beta \h V_{d-1}}=e^{{2d \over d-2}\phi_d} \h \cf$. Therefore the Einstein frame action is 
\begin{align}
	S=\int d^dx \sqrt{-g}\, \left[ \left({ \R \over 2}-{2\over d-2}(\partial \phi_d)^2+\dots\right)- \cf\left(\beta\, ,\phi_d, \dots\right) \right]. \label{EffAct1LEin}
\end{align}
Effectively this action describs a universe filled with a fluid of free energy density $\cf$, and this cosmic fluid is just the ideal string gas. 
We then look for cosmological solution, which, by virtue of the perturbative approach, is meant to be a perturbative correction added to the tree level flat spacetime. Thus we use the ansatz of flat FRW metric for the solution at one-loop order, which is written down in the Einstein frame as
\begin{align}
	d s^2=-N(x^0)^2 (dx^0)^2+\sum_{i=1}^{d-1}a(x^0)^2 (dx^i)^2,\ \ ( 0\leq x^i <1). \label{MetAnz}
\end{align}
Here $x^0$ is the coordinate time, $x^i$ ($i=1,\dots,d-1$) are spatial coordinates as in Eq.(\ref{THMetric}), $a(x^0)=2\pi \h R_{\rm box}(x^0) \,e^{{2\over d-2}\phi_d}$ is the scale factor, and $N(x^0)=\h \beta(x^0) \,e^{{2\over d-2}\phi_d}=\beta(x^0)$ is the lapse function identified here with the Einstein frame inverse temperature. These quantities, as well as the scalar fields, which are static at tree level, can now have nontrivial time dependence due to the inclusion of one-loop thermal and quantum effects.
Moreover these time evolutions, which are meant to divert only perturbatively from the tree-level static solution as is stated above Eq.(\ref{MetAnz}), should be quasi-static. Otherwise if the metric and the fields evolve violently, kinetic corrections become important and should be included in the action, and thus Eq.(\ref{EffAct1LEin}) does not apply any more.

Now we are ready to derive the equations of motion. For simplicity we switch off all scalar fields but the dilaton, while it is straightforward to include more scalar fields. Plugging Eq.(\ref{MetAnz}) into the effective action (\ref{EffAct1LEin}), varying with respect to $N$, $a$ and $\phi_d$, we obtain
\begin{align}
    \demi (d-1)(d-2)  H^2 - {2\over d-2}\dot \phi_d^2  &=\beta {\partial \cf \over \partial \beta}+\cf\, , \label{EOM1} \\
    \demi (d-2)\,\Big[2\dot H+(d-1) H^2 \Big] +{2\over d-2}\dot \phi_d^2 &= {a\over d-1}{\partial \cf \over \partial a}+ \cf\, , \label{EOM1bis}\\
    \ddot \phi_d  + (d-1)H\, \dot\phi_d + {d-2 \over 4}{\partial \cf \over \partial \phi_d} &=0, \label{EOM2} 
\end{align}
where we assumed the dilaton, as well as other scalar fields if they were included, is spatially homogeneous. Also $H=\dot a /a$ is the Hubble parameter, and the dot stands for the derivative with respect to the cosmic time $t=t(x^0)$ defined by $dt=N(x^0)\, dx^0$. We use $t$ in the equations of motion in order to absorb $N$ into the derivatives to simplify the equations. If there were more scalar fields, then their equations of motion, which is of a similar form of Eq.(\ref{EOM2}), should be added to the above list, and also their kinetic terms should appear in parallel with that of the dilaton in the first two equations (\ref{EOM1}) and (\ref{EOM1bis}). Referring to the standard form of Friedemann equations the right-hand-side of Eqs (\ref{EOM1}) and (\ref{EOM1bis}) should be the energy density $\rho$ and minus the pressure $-P$ of the cosmic fluid. Therefore we read off the energy density and the pressure of the string gas from the cosmological equations of motion:
\begin{align}
	\rho=\beta {\partial \cf \over \partial \beta}+\cf, \hspace{5mm}P=-{a\over d-1}{\partial \cf \over \partial a}-\cf, \label{rhoP}
\end{align}
which are measured in the Einstein frame. Then using the relation $Z=-\beta V_{d-1} \cf =-\beta F$ and $V_{d-1}= a^{d-1} $, we rewrite the above expressions as
\begin{align}
	\rho=-{\partial Z\over \partial \beta}\, , \hspace{5mm}P=-{\partial F \ \ \over \partial V_{d-1}}\, ,
\end{align}
which are exactly the standard canonical ensemble relations. Therefore the cosmological $\rho$ and $P$ in Eq.(\ref{rhoP}) are really the energy density and pressure in the thermodynamical sense, as long as the string gas is constantly in equilibrium during the cosmological evolution. This is actually the case, guaranteed by the quasi-static evolution of the universe, stated below Eq.(\ref{MetAnz}). There is one last equation of importance: the continuity equation, which can be derived from Eqs (\ref{EOM1})---(\ref{EOM2}):
\begin{align}    
        \dot \rho+(d-1)H(\rho+P) = \dot\phi_d \, \, {\partial \cf\over \partial \phi_d}\, . &\label{EOM3}
\end{align}
In case where more scalar fields are present, the right hand side should be $\sum_{\Phi} \dot \Phi \left(\partial \cf / \partial \Phi\right)$ where $\Phi$ runs over all scalar fields. The continuity equation can be integrated, using $\dot \cf=\dot \beta {\partial \cf \over \partial \beta}+ \dot \phi_d{\partial \cf\over \partial \phi_d}+(\text{terms with other scalars})$, to give
\begin{align}    
        {d\over dt}\left[ \beta\, a^{d-1} (\rho+P)\right]=0 \ \ \Leftrightarrow\ \    \beta\, a^{d-1} (\rho+P)={\rm const}. &\label{ConEnt}
\end{align}
which is just the conservation of entropy, given that $a^{d-1}=V_{d-1}$ so that $\beta a^{d-1} (\rho+P)=\beta(E+PV_{d-1})=S$. This equation holds regardless of the content of scalar fields, and can be used in place of Eq.(\ref{EOM1bis}).

We recap the essential points in the construction that we have gone through: 

\noindent $\bullet$ We examine a critical and weakly coupled string theory at finite temperature, with the aim of looking for solutions corresponding to cosmological evolution. 

\noindent $\bullet$ The search for cosmological solution at tree level is unsuccessful, so that we go to one-loop level, where the one-loop correction to the effective action is computed against a thermal background.

\noindent $\bullet$ We end up with an effective theory describing a universe filled with an ideal string gas in thermal equilibrium, and the effective theory action holds for a universe in quasi-static evolution.


\subsection{Example: attraction to a radiation-dominated universe} \label{EOMEX}


%

%
%


Now we would like to show a concrete example based on the string theory model in Sec.\ref{9dim}: the heterotic string in 9 dimensions. Certainly it is phenomenologically more relevant to consider 4 dimensional models where the extra dimensions compact of order string length. 
However we consider here 9 dimensional cosmology for simplicity, while technically reducing from 9 dimensions to 4 dimensions amounts to enlarging the moduli space and including more fields in the effective action without changing the form of the equations of motion.

The thermal background space at tree level is
\begin{align}
	S^1(\h R_0)\times T^8(\h R_{\rm box}) \times S^1(R_9).
\end{align}
Here $R_9$ is indifferent of reference frame so we do not put hat on it. The one-loop amplitude against this background is already worked out in Eq.(\ref{Unf9D2}), in terms of string frame quantities. 
The effective action is
\begin{align}
\label{stab81}
    S =\int d^9x\sqrt{-g}\, \left[\frac{\R}{2}-\frac{2}{7}\,  \partial \phi_9 \partial \phi_9 - \demi \partial \Phi  \partial \Phi -\F(\beta, \phi_9,\Phi) \right]
\end{align}
where we have defined $\Phi:=\ln R_9$ for notational simplicity, since $\Phi$ has canonical kinetic term. Varying Eq.(\ref{stab81}), or using directly the general equations (\ref{EOM1}) to (\ref{ConEnt}), we obtain the set of independent equations of motion
\begin{align}
	& 28\, H^2=\frac{2}{7}\,  \dot{\phi}_9^2+\demi\,  \dot{\Phi}^2  +\rho,\label{eomR1}\\
    & \ddot{\phi}_9 + 8H\dot{\phi}_9 + \frac{7}{4}\, {\partial \cf \over \partial \phi_9}=0, \label{eomR3} \\ \label{eomR4}
    &\ddot \Phi +8 H\dot\Phi + {\partial \cf \over \partial \Phi}=0,\\ \label{eomR5}
    & \beta\, a^8  (\rho+ P)={\rm const.} \, ,
\end{align}
To further narrow down the discussion, we enforce some extra restrictions: {\em i)} we assume the background scalar fields are constant, or small fluctuations about some constant values; {\em ii)} the string frame temperature is low with respect to the string scale: $\h R_0 \gg 1$, which allows us to use directly the result (\ref{FreeEn}) for $\cf$; {\em iii)} the internal radius $R_9$ is restricted in the vicinity of 1, so that we focus our attention to phase I in Fig.\ref{ptHet}, corresponding to the first line in Eq.(\ref{FreeEn}).

We look for solutions with static scalar background $\dot \phi_9=\dot \Phi=0$ which means $\partial_{\phi_9} \cf=\partial_{\Phi} \cf=0$. Restricted to the phase I of Fig.\ref{ptHet}, we see that this is possible only at the local minimum $R_9=1$ or $\Phi=0$. At this point the dilaton $\phi_9$ becomes a flat direction since $\cf$ does not depend on it. Thus we have, referring to the first line of Eq.(\ref{FreeEn}),
\begin{align}
	&\phi_9={\rm const.} \ \ {\rm such\ that }\ \ e^{\phi_9}\ll 1; \ \ \ \Phi=0; \label{phiST} \\
	&\cf=- \left(\S_0\T_0+2\S_0 \T_{-1}\right) G_9(0) \, T^9, \label{cfST}
\end{align}
where $T=\beta^{-1}$ is the Einstein frame temperature. From Eq.(\ref{cfST}) we derive the energy density and the pressure
\begin{align}
	P={1\over 8}\, \rho=-\cf =  \left(\S_0\T_0+2\S_0 \T_{-1}\right) G_9(0)\, T^9,
\end{align}
which is the Stefan's law for radiation in 9 dimensions. Plugging $\rho$ and $P$ in Eqs (\ref{eomR1}) and (\ref{eomR4}), we obtain the following cosmological evolution which describes a 9-dimensional radiation-dominated universe
\begin{align}
	a(t)\propto {1\over T(t)} \propto t^{2/9}. \label{RDS}
\end{align}
We do not intend to show the calculation here but this solution is stable against small perturbations\cite{Estes:2011iw}. As a result, the scalar field $\Phi=\ln\! R_9$, massless at tree level, obtains a mass at one-loop level at $\Phi=0$, the local minimum of $\cf$. The mass squared is given by the second derivative of Eq.(\ref{FreeEn}) with respect to $\Phi$:
\begin{align}
	M_{\Phi}^2=\left.{\partial^2\cf \over \partial \Phi^2}\right|_{\Phi=0}={4 \over \pi} \, \S_0\T_{-1}G_{7}(0) \, e^{{4\over 7}\phi_9}\, T^7, \label{eq86}
\end{align}
where the factor $e^{{4\over 7}\phi_9}$ appears because we work in Einstein frame. This gives an example of moduli stabilization by thermal effects, where a remarkable feature, which is generically true of such moduli stabilization is that the induced scalar mass is temperature-dependent, and is therefore time-dependent due to Eq.(\ref{RDS}).
 
Despite the awkward dimensionality 9, the following features of this model persist in other dimensions higher than or equal 4, as long as the temperature is well below the Hagedorn temperature \cite{stabmod2,Estes:2011iw}:

\noindent $\bullet$ The cosmological solution converges with cosmic time $t$ to an evolution according to $a(t)\propto T(t)^{-1}\propto t^{2/d}$, the behavior of a radiation-dominated universe. 

\noindent $\bullet$ The solution is stable against fluctuations in the scalar fields, showing that it is possible to stabilize moduli. That is, scalar fields massless at tree level can obtain masses at one-loop level at local minima of the one-loop free energy density.

\noindent $\bullet$ The dilaton is frozen at a fixed value (can decrease logarithmically in 4 dimensions but can never increase \cite{Estes:2011iw}), so that the model remains in the weak coupling regime once it starts out weakly coupled.

%
%
%
%
%

\subsection{Other issues and perspectives}

Despite the radiation-dominated early universe that it gives rise to, the example in the last subsection is of little phenomenological interest, obviously because it misses many pieces for completing the jigsaw puzzle of the real universe. For one thing, the solution (\ref{RDS}) is valid for temperature well below the Hagedorn temperature, since otherwise we loose analytic control of thermal and quantum corrections, and we cannot use the results in Sec.\ref{TISG}. On the other hand no explicit matter formation process is considered. Thus the solution (\ref{RDS}) should correspond to a short time span between the exit of the Hagedorn phase and the electoweak symmetry breaking, not counting the incorrect spacetime dimensionality, and the unbroken spacetime supersymmetry... In this very last part of our lecture notes, we briefly discuss how thermal string cosmology handle these issues.


\subsubsection*{Hagedorn phase and the initial cosmological singularity}


Due to the lack of an analytic description at full string level, in our thermal string scenario it is not yet possible to work out the cosmology in the Hagedorn phase. However there have been some fully solvable models of ideal string gas constructed, in which the Hagedorn temperature does not arise, and which can lead to the resolution of the initial cosmological singularity.

The first such model was constructed in \cite{Angelantonj:2008fz}, based on perturbative type II strings. Switching on fluxes along the Euclidean time circle and another internal direction in a certain way, one can deform the mass spectrum such that the thermal winding modes responsible for the Hagedorn instability can no longer become tachyonic. Instead they only become massless at a critical temperature $\beta_c$ of about the string scale, and $\beta_c^{-1}$ is the maximal temperature of the string gas, at which the thermodynamic quantities are finite. Also the model is invariant under the thermal T-duality $\beta\rightarrow \beta_c^2/\beta$.

Later on, by the same construction a 2 dimensional model free of Hagedorn temperature was realized \cite{Florakis:2010ty} and its cosmology worked out \cite{Florakis:2010is}. It was found that if we let the expanding universe evolve backwards, the temperature raises gradually with the quasi-static contraction of the universe until its maximal value $\beta_c^{-1}$. There some thermal winding modes become massless, inducing a gauge symmetry enhancement in the Euclidean time direction $U(1)\rightarrow SU(2)$. These massless  thermal winding modes source a space-like brane of positive tension, which supplies a sudden negative pressure preventing the universe from further contraction. The universe then embarks on a quasi-static expanding phase, and therefore we obtain a quasi-static bouncing universe. The temperature and the dilaton are at their maximum values exactly at the bouncing point, and drop when the universe evolves forwards or backwards. The behavior of the dilaton shows that the model can be maintained in the weakly coupled regime, which, together with the quasi-static evolution, ensures that the perturbative calculations are valid. 

As the generalization of \cite{Florakis:2010is}, quasi-static bouncing universes were achieved in \cite{Kounnas:2011fk} in arbitrary spacetime dimensions via the same mechanism. Then in \cite{Kounnas:2011gz} the authors studied the case where the space-like brane exists for an extended time span, where bouncing cosmological solutions were also found.


\subsubsection*{Moduli stabilization}


The example shown in Sec.\ref{EOMEX} has already touched upon this issue. Being the simplest case of compactification, the moduli space is two dimensional, coordinatized by the dilaton $\phi_9$ and the internal radius $R_9$. 
These flat directions are corrected at one-loop level via the free energy density Eq.(\ref{FreeEn}). As a result the internal radius is stabilized at the point $R_9=1$ of enhanced gauge symmetry $U(1)_{R_9} \rightarrow SU(2)_{R_9}$, while the flat direction $\phi_9$ is not lifted. 

The conception of such mechanism goes back to \cite{GVNSS}, where it was shown, for non supersymmetric compactifications of heterotic string, that the cosmological constant generated by string loops reaches extremum at points of enhanced gauge symmetry. This feature was proposed in \cite{Angelantonj:2006ut} for use of moduli stabilization where moduli are attracted to the local minima of the moduli dependent cosmological constant generated perturbatively beyond tree level. In the context of thermal string cosmology this mechanism for moduli stabilization is extensively investigated in various models. The works in \cite{cosmoB,stabmod2,cosmoreview} have addressed the moduli stabilization by perturbative effects at one-loop level, while non-perturbative effects are studied in \cite{Estes:2011iw,Liu:2011nw} via string theory dualities.




The cosmological moduli problem should also be taken into account. With the cosmological evolution, the oscillations of the stabilized moduli in their potential wells store an energy density which dominates over that of the radiation \cite{Preskill:1982cy}. This leads to a huge entropy production when the scalar particles of the moduli fields decay, which wipes out the baryon-antibaryon asymmetry. Also the production of the decay can alter the primordial abundances of light nuclei produced by nucleosynthesis, leading to contradictions with observation \cite{CosmModProb}. Usually the problem is solved by imposing a lower bound of about ${\cal O}(10)$TeV to the induced scalar mass of the stabilized moduli \cite{Ellis:1986zt}, for example in the KKLT senario \cite{Kachru:2003aw}. In the thermal string cosmology scenario however, the induced scalar mass is time dependent, as is inferred from Eq.(\ref{eq86}), based on which one can show that the energy stored in the moduli field oscillation never exceeds that of the radiation \cite{stabmod2,cosmoreview,Estes:2011iw}. As a result, the cosmological moduli problem does not arise.

\subsubsection*{Supersymmetry breaking and hierarchy problem}

In supergravity unlike in supersymmetric field theory, at tree level it is possible to have spontaneous supersymmetry breaking with the cosmological constant still being zero. Such models are known as the no-scale supergravity \cite{Noscale}, which are of particular interest in phenomenological applications. 

In superstring theory, a wide class of orbifold compactifications which spontaneously break spacetime supersymmetry can lead to the no-scale supergravity as the tree-level effective theory (see \cite{Kounnas:2007hb} for a summary and references), among which the Scherk-Schwarz compactifications have been intensively exploited in thermal string cosmology. 
The vanishing tree-level cosmological constant implies that the tree-level vacuum is still given by Eq.(\ref{BG1}), so that the technique depicted in Sec.\ref{TISG} can still be applied when we go to the one-loop level. The behavior of the supersymmetry breaking scale can be precisely traced, since it is one of the moduli fields (the no-scale modulus) and thus its equation of motion can be explicitly written down.

Quantitative computations were carried out in \cite{cosmoC,stabmod1,stabmod2,cosmoreview}, where heterotic string theory is compactified on orbifolds down to four dimensions with supersymmetry spontaneously broken \`a la Scherk-Schwarz, of the pattern $\cn_4=2\rightarrow 0$ \cite{cosmoC} and $\cn_4=1 \rightarrow 0$ \cite{stabmod1}. It was found that the resulting cosmological behavior is sensitive to the details of the compactification, where the most interesting situation is where the no-scale modulus, or the supersymmetry breaking scale, say $M$, decreases proportionally with temperature, and where the scale factor evolves as that of a radiation-dominated universe (referred to as radiation-like universe in the original work): $M(t)\propto T(t) \propto a(t)^{-1}\propto t^{-1/2}$ with $t$ the cosmic time. We thus achieve in such universes the dynamical generation of the hierarchy $M\ll M_{\rm Planck}$. Also the dilaton decreases with cosmic time, which maintains the validity of the perturbative calculations.

\subsubsection*{Radiative electroweak symmetry breaking}

The universe as illustrated in the previous paragraph runs into trouble at late time, due to the proportionality between the supersymmetry breaking scale $M(t)$ and the temperature $T(t)$. As the temperature decreases to almost zero, so does the supersymmetry breaking scale, and therefore the universe will end up supersymmetric. However there can be mechanism in certain models which can break the proportionality bind $M(t)\propto T(t)$ and halts the evolution of $M(t)$ before it gets too small. Radiative symmetry breaking by IR effects is the candidate mechanism, which is well understood in the context of supergravity \cite{AlvarezGaume:1983gj}. Such mechanism defines a transmutation scale $Q$ of about the electroweak scale. As the energy of the system, here characterized by temperature, drops below $Q$, radiative corrections starts to destabilize the scalar potential and triggers the electroweak phase transition. Hopefully this is accompanied with the stabilization of $M$, while temperature continues to decrease. It will be interesting to work out the details of this mechanism.




\subsubsection*{Primordial cosmological fluctuation}


A further direction to explore is to make connection with the observational data of the cosmic microwave background in the context of the bouncing cosmology in \cite{Florakis:2010is,Kounnas:2011fk,Kounnas:2011gz}. In light of the matter-bouncing scenario \cite{Finelli:2001sr}, where it is shown that a nearly scale invariant spectrum of cosmological fluctuations can be produced, work is in progress 
studying the propagation of cosmological fluctuations across the space-like brane from the contracting phase to the expanding phase \cite{Kounnas:2013yda}. If successful, this can eventually provide an alternative scenario to that of the inflationary cosmology. We would like to notify in the end that the same problem has been addressed recently in string gas cosmology \cite{Brandenberger:2008nx}, and also there have been large amount of work on string cosmology in favor of inflation \cite{McAllister:2007bg} which is still an active field.

\subsection*{Acknowledgments}

I am especially grateful to the organizers of the Modave Summer School VIII for the invitation to give these lectures. I also thank the Modave Summer School attendees for their feedbacks, which are important to the preparation of these lecture notes. Many thanks to Herv\'e Partouche who proofread the manuscript and gave helpful comments and suggestions. Part of this work was done at the Centre de Physique Th\'eorique, Ecole Polytechnique, supported by the EU contracts PITN GA-2009-237920, ERC-AG-226371 and IRSES-UNIFY, the French ANR 05-BLAN-NT09-573739 contract, the CEFIPRA/IFCPAR 4104-2 project, and PICS contracts France/Cyprus, France/Greece and France/US. The part of this work accomplished at the Institute for Theoretical Physics, KULeuven, is supported by a grant of the John Templeton Foundation.

\appendix

\vspace{1cm}

\section{Canonical partition function in quantum field theory at one-loop} \label{app1}

This appendix reviews the standard computation in finite temperature quantum field theory for obtaining the canonical partition function of an ideal gas of particles. The results are used in Sec.\ref{NonThAmpSec} and Sec.\ref{TISG} for motivating the non-thermal and thermal formalisms in string theory.

We start with the field theory action describing one particle degree of freedom of mass $M$ in a flat $d$-dimensional Euclidean space, which is
\begin{align}
    S=\int d^d x \,\phi\, \big(\!  -\DAL_{\rm E}+M^2\big)\, \phi, \label{LA2Q1}
\end{align}
where $\DAL_{\rm E}=\partial_0^2+\vec \nabla^2$ is the $d$-dimensional Laplacian. 
The canonical partition function $\cz = {\rm Tr}\, e^{-\beta H}$ can be evaluated by the path integral
\begin{align}
    \ln \cz_{\rm B,  F} =&\ln\Big(\int\cd \phi\,\, e^{-S[\phi]}\Big)=\ln\Big[{\rm Det}\big(- \DAL_{\rm E}+M^2\big)\Big]^{\mp 1/2} = \mp\demi {\rm Tr} \ln\big(\! -\DAL_{\rm E}+M^2 \big), \label{LA2Q2}
\end{align}
where the Euclidean time is compactified on a circle $S^1(R_0)$ of radius $R_0=\beta/2\pi$, with $\beta$ the inverse temperature. The subscripts ``B'' and ``F'' stand for boson and fermion, corresponding respectively to the signs ``$-$'' and ``$+$'' in the second and the third step. Here for simplicity we have cheated in the fermionic case, computing the path integral as if $\phi$ were a real grassmanian variable and $- \, \DAL_{\rm E}+M^2$ were an antisymmetric matrix. The result is correct which is not accidental since each single fermionic degree of freedom satisfies the Klein-Gordon equation, the equation of motion derived from the action (\ref{LA2Q1}). 

Eq.(\ref{LA2Q2}) is nothing but the one-loop amplitude of the field $\phi$. In evaluating Eq.(\ref{LA2Q2}), $\phi$ takes periodic boundary condition along $S^1(R_0)$ for a boson and anti-periodic boundary condition along $S^1(R_0)$ for a fermion, leading to the following eigenvalues for the Laplacian operator $\DAL_{\rm E}$:
\begin{align}
	 -\left({2\pi \, r\over R_0}\right)^2- \vec p^{\, 2}\,, \ \ \ \ \ \  \ r\in \left\{ \begin{array}{ll} \mathbb Z & \text{boson} , \\   \mathbb Z+\demi & \text{fermion}, \end{array} \right. \label{rrrr}
\end{align}
where $\vec p$ is the $(d-1)$-dimensional spatial momentum in the non-compact directions which takes continuous values. The discrete momentum modes in the Euclidean time direction are the so called Matsubara modes.

We can evaluate Eq.(\ref{LA2Q2}) using the integration representation of the logarithm
\begin{align}\label{form7}
    \ln A=-\int_0^{\infty}\frac{d\ell}{\ell}\big( e^{-A \, \ell}-e^{-\ell} \big)\, .
\end{align}
For one bosonic degree of freedom, we have
\begin{align}
    \ln\cz_{\rm B}= &-\demi {\rm Tr}\, \ln\big(\! -\DAL_{\text{E}}+M^2 \big) =\demi {\rm Tr} \, \int_0^{\infty}\frac{d \ell}{\ell}\lbk e^{-\ell \, \left(-\, \DAL_{\text{E}}+M^2 \right)} -e^{- \ell}\rbk\nnR \label{C4}
     =&{V_{D-1}\over 2(2\pi)^{D-1}}\sum_{m\in \mathbb Z}\int d \vec p \int_0^{\infty}\frac{d\ell}{\ell}\, \bigg[ e^{-\ell \,\left(\frac{m^2}{R_0^2}+\vec{p}^2+M^2 \right)}- e^{- \ell}\bigg].
\end{align}
In this expression $\ell$ is referred to as the Schwinger time parameter, whose physical interpretation is the proper time that the particle spends for completing the loop.\footnote{The fact that the logarithm of the partition function is a one-loop amplitude is obscure in the calculation presented so far in this section. A more detailed analysis showing $\ln \cz =\begin{minipage}{6mm} \includegraphics[width=6mm]{circle_no.pdf} \end{minipage}$ is presented in the Appendix A of \cite{Liu:2012yka}.} The second term in the bracket in Eq.(\ref{C4}) has no temperature dependence, which is thus an irrelevant infinity and can be discarded naively. In fact the Riemann-Zeta regularization of the sum over $m$ also yields zero. Then we integrate over the spatial momentum $\vec{p}$ which is just a $(d-1)$-dimensional Gaussian integration, and perform Poisson resummation over the Matsubara mode index $m$ using 
\begin{align}\label{form9}
	\sum_{\tilde{m}}e^{-\pi\tilde{m}^{\text{T}}A\tilde{m}+2\pi i b^{\text{T}}\tilde{m}}
	=\frac{1}{\sqrt{\text{det}A}}\sum_{m}e^{-\pi(m-b)^{\text{T}}A^{-1}(m-b)},
\end{align}
and we get\footnote{The power of the Schwinger parameter $-(1+d/2)$ is characteristic of one-loop amplitudes, different to $-d/2$ in the case of propagators, where the topology of the particle trajectory is a segment (c.f. for example Eq.(2.1.28) in \cite{Kiritsis:2007zza}). The extra $-1$ in the one-loop amplitude is for removing the redundancy of the rigid global rotation of the loop, which does not exist for a segment with fixed end points.}
\begin{align}\label{C5}
    \ln\cz_{\rm B} = &\, \frac{\beta V_{d-1}}{2(4\pi)^{d/2}}\sum_{\t m}\int_0^{\infty} \frac{d\ell}{\ell^{1+d/2}}\, \exp\Big(\! -\frac{\pi^2R_0^2}{\ell }\, \t m^2-\ell M^2 \Big)\nnR =& \, \frac{\beta V_{d-1}}{2(2\pi)^d}\sum_{\t m}\int_0^{\infty} \frac{d\ell}{\ell^{1+d/2}}\, \exp\Big(\! -\frac{\pi R_0^2}{\ell }\, \t m^2-\pi \ell M^2 \Big),
\end{align}
where in the second line we have simply recaled the Schwinger parameter, so that the expression is more adapted to string theory language. Also with Poisson resummation we trade the the sum over Matsubara modes $m$ in Eq.(\ref{C4}) for the sum over instanton number $\t m$ in Eq.(\ref{C5}), where the instanton configuration labeled by $\t m$ is such that the particle worldline winds around the Euclidean time circle $\t m$ times before completing the loop. For this reason, here we call these instantons {\em worldline instanton}.

The case of one fermionic degree of freedom is calculated similarly:
\begin{align}
    \ln\cz_{\rm F}= &\, \demi {\rm Tr}\, \ln\big(\! -\, \DAL_{\text{E}}+M^2 \big) =- \demi {\rm Tr} \, \int_0^{\infty}\frac{d \ell}{\ell}\lbk e^{-\ell (-\, \DAL_{\text{E}}+M^2)} -e^{- \ell}\rbk\nnR\label{C6}
     =& \,-\frac{V_{D-1}}{2(2\pi)^{D-1}}\sum_{m}\int d \vec p \int_0^{\infty}\frac{d\ell}{\ell}\, \bigg\{ e^{-\ell \, \left[\frac{(m+1/2)^2}{R_0^2}+\vec{p}^2+M^2 \right]}- e^{- \ell}\bigg\},
\end{align}
where the Matsubara modes $m$ are shifted by $1/2$ due to the anti-periodic boundary condition as is shown in Eq.(\ref{rrrr}). Now performing Poisson resummation over $m$, integrating over $\vec p$, and removing the physically irrelevant infinity, we arrive at
\begin{align}
    \ln\cz_{\rm F} =\, &\frac{\beta V_{d-1}}{2(4\pi)^{d/2}}\sum_{\t m}\int_0^{\infty} \frac{d\ell}{\ell^{1+d/2}}\, (-1)^{\t m+1} \exp\Big(\! -\frac{\pi^2R_0^2}{\ell }\, \t m^2-\ell M^2 \Big)\nnR =\, &- \frac{\beta V_{d-1}}{2(2\pi)^d}\sum_{\t m} \int_0^{\infty} \frac{d\ell}{\ell^{1+d/2}}\, (-1)^{\t m} \exp\Big(\! -\frac{\pi R_0^2}{\ell }\, \t m^2-\pi \ell M^2 \Big).\label{C7}
\end{align}
Here different from the bosonic case Eq.(\ref{C5}), the instanton sum is weighted by an interchanging sign $(-1)^{\t m}$, accounting for the statistics of fermions.

It will be useful to consider the case where the ideal gas consists of several species of particles. In this case the total partition function should be the product of the partition functions of each species, since the total Hilbert space is a direct product of the component Hilbert spaces. Therefore the logarithm of the total partition function is merely an addition of the logarithm of component partition functions. We assume the compound ideal gas has mass spectrum $\{M_s\}$ with helicities $\{j_s\}$, where $s$ is the index of quantum states, and we have
\begin{align}
	\ln \cz= \sum_s \sum_{\t m} \frac{\beta V_{d-1}}{2(2\pi)^d} \int_0^{\infty} \frac{d\ell}{\ell^{1+d/2}}\, (-1)^{2 j_s(\t m+1)} \exp\Big(\! -\frac{\pi R_0^2}{\ell }\, \t  m^2-\pi \ell M_s^2 \Big)\, . \label{lnZtot}
\end{align}

It is instructive to make connection with the standard canonical ensemble formalism. To this end we make use of the following mathematical formulae
\begin{align}
	&\int_0^{\infty} {dt }\, \, t^{\, -\nu}\, e^{-{A \over t}-B\, t} =2\left({B \over A}\right)^{\nu-1\over 2} K_{\nu-1}\left(2\sqrt{AB}\right), \ \ {\rm for}\ {\rm Re}A\ {\rm and}\ {\rm Re}B\ {\rm positive}; \label{Knu} \\
    &\int_0^{\infty}p^n dp\, \, e^{-\beta \sqrt{p^2+M^2}}={2^{n\over 2}\over \sqrt \pi}\, \Gamma\Big({n+1 \over 2}\Big)\, M \Big({M \over \beta}\Big)^{n\over 2} K_{n+2 \over 2}(\beta M), \label{FMLA8}
\end{align} 
where $K_{\nu}(\, \cdot\,)$ is the modified Bessel function of the second kind. With the above two relations as well as the Taylor expansion of the logarithmic function, we can actually show that
\begin{align}
	{\beta \over 2\pi} \sum_{\t m=1}^{\infty}\int_0^{\infty} \! {d\ell\over \ell^{1+d/2}} \, \exp\Big(\! -\frac{\pi R_0^2}{\ell }\, \t  m^2-\pi \ell M^2 \Big) = -\int \! d\vec p \, \ln\left(1-e^{-\beta \sqrt{\vec p^2+M^2}}\right), \\
	{\beta \over 2\pi} \sum_{\t m=1}^{\infty}\int_0^{\infty} \! {d\ell\over \ell^{1+d/2}} \, (-1)^{\t m+1}\exp\Big(\! -\frac{\pi R_0^2}{\ell }\, \t  m^2-\pi \ell M^2 \Big) = \int \! d\vec p \, \ln\left(1+e^{-\beta \sqrt{\vec p^2+M^2}}\right), 
\end{align}
where $\vec p$ denotes the $(d-1)$-dimensional spatial momentum, and $M$ is some mass. We can now rewrite Eq.(\ref{lnZtot}) as
\begin{align}
	\ln \cz= & \sum_s  \frac{\beta V_{d-1}}{2(2\pi)^d} \int_0^{\infty} \frac{d\ell}{\ell^{1+d/2}}\, (-1)^{2 j_s} e^{ -\pi \ell M_s^2 } \nnR &\ \ \ \ \ \ - \sum_s {V_{d-1}\over (2\pi)^{d-1}} \int\! d\vec p\, (-1)^{2j_s}  \ln \left[1- (-1)^{2j_s}e^{-\beta \sqrt{\vec p^2+M_s^2}}\right]\, .\label{lnZstand}
\end{align}
Thus we recover the standard canonical ensemble formalism, where the first line is merely the Casimir energy.

Finally we mention another case of our interest, which is the non-thermal limit of Eq.(\ref{lnZtot}). At this limit the Euclidean time decompactifies $R_0\rightarrow \infty$, and all the instanton configurations are exponentially suppressed in Eq.(\ref{lnZtot}) except for $\t m=0$, and also we replace $\beta V_{d-1}$ by $V_d$ the total spacetime volume. Thus we obtain
\begin{align}
	\ln \cz\big|_{T=0} = \sum_s  \frac{ V_d }{2(2\pi)^d} \int_0^{\infty} \frac{d\ell}{\ell^{1+d/2}}\, (-1)^{2 j_s} \exp\Big(\! -\pi \ell M_s^2 \Big)\, . \label{lnZtotT0}
\end{align}
The same result can be obtained putting $\beta\rightarrow \infty$ in Eq.(\ref{lnZstand}), where the second line vanishes.

\section{Dedekind $\eta$-function, Jacobi elliptic functions and $SO(8)$ characters} \label{tttttt}

In this appendix we give a pragmatic summary of the important functions and notations used in Sec.\ref{ExNonTh} in the string one-loop amplitudes. The Appendix C of \cite{Kiritsis:2007zza} is recommended for more details. Some detailed explanation about how these functions actually arise in the computation of one-loop amplitudes can be found in \cite{Angelantonj:2002ct,Lust:1989tj,Ginsparg:1988ui}. The functions that we need are
\begin{align}
	\text{Dedekind $\eta$-function:   }  &\eta(\tau)=\, q^{1/24}\prod_{n=1}^{\infty} (1-q^n), \label{DefEta} \\
	\text{Jacobi elliptic function:   }&\theta[^a_b](\tau)= \sum_{k\in \mathbb Z} q^{\demi \left(k-{a\over 2} \right)^2} e^{-i\pi\,b \, \left(k-{a\over 2} \right)}, \label{DefTta}
\end{align}
where $q=e^{2\pi i \tau}$, with $\tau$ sitting in the upper-half complex plane. For our purpose we will only encounter cases where $a, b=0,1 {\rm\, mod\, }2$. We will also use the complex conjugates $\b \eta$ and $\b \theta[^a_b]$, which are obtained by replacing in the above definitions $q$ with $\bar q=e^{-2\pi i \b \tau}$. 

When we want to write down the string one-loop amplitudes, it is convenient to work in the lightcone gauge, where the general rules are: {\em each holomorphic (anti-holomorphic) worldsheet boson contributes a factor of $\eta^{-1}$ ($\b \eta^{-1}$), and each holomorphic (anti-holomorphic) worldsheet fermion contributes a factor of $\sqrt{\eta^{-1} \theta[^a_b]}$ ($\sqrt{\b \eta^{-1} \b \theta[^a_b]}\, $)} where $a$, $b$ take $0$ or $1$ independently, depending on the spin structure of the worldsheet fermions on the torus (cf for example \cite{Ginsparg:1988ui}). 

It is useful to further define $SO(8)$ characters out of the $\eta$- and $\theta$-functions, with which the expression of one-loop amplitudes can better reveal the structure of the superstring spectra. Here it is $SO(8)$ in particular because the lightcone gauge fixing of superstring theories leaves us with 8 target space dimensions. As a result the spectrum of spacetime states fall into different representations of $SO(8)$.

The $SO(8)$ characters which we use are \cite{Bianchi:1990yu,Angelantonj:2002ct}
\begin{align}
	O_8= &{\theta[^0_0]^4 +\theta[^0_1]^4 \over 2 \eta^4} =q^{-{1\over 6}} \, \big[1+28q+134\, q^2+568\, q^3+ {\cal O}(q^4)\big], \label{SO8chO} \\
	V_8= &{\theta[^0_0]^4 -\theta[^0_1]^4 \over 2 \eta^4} =q^{1\over 3} \, \big[8+64\, q+288\, q^2+1024\, q^3+{\cal O}(q^4)\big], \label{SO8chV} \\
	S_8= &{\theta[^1_0]^4 +\theta[^1_1]^4 \over 2 \eta^4} =q^{1\over 3} \, \big[8+64\, q+288\, q^2+1024\, q^3+{\cal O}(q^4)\big], \label{SO8chS} \\
	C_8= &{\theta[^1_0]^4 -\theta[^1_1]^4 \over 2 \eta^4} =q^{1\over 3} \, \big[8+64\, q+288\, q^2+1024\, q^3+{\cal O}(q^4) \big], \label{SO8chC}
\end{align}
which arise from the sectors of the superstring spectrum whose ground states transform in the scalar ($O_8$), vector ($V_8$) and two real chiral spinor ($S_8$ and $C_8$) representations of $SO(8)$. These symbols can also appear as $\chi_{{}_O}$, $\chi_{{}_V}$, $\chi_{{}_S}$, $\chi_{{}_C}$ in the literature, for example in \cite{Kiritsis:2007zza}. We will equally use the complex conjugates $\b O_8$, $\b V_8$, $\b S_8$ and $\b C_8$, obtained by replacing $q$ with $\b q$ in the above definitions. Upon quantization of worldsheet fermions in superstring theories, the Neveu-Schwarz sector states give rise to $V_8$, while the Ramond sector states give rise to $S_8$ or $C_8$. 

The following identities will be useful
\begin{align}
\demi \sum_{a,b=0}^1(-1)^{a+b+ab}\, {\theta[^a_b]^4 \over \eta^4}=V_8-S_8,\ \ \ \demi\sum_{a,b=0}^1(-1)^{a+b} \, \ {\theta[^a_b]^4 \over \eta^4} =V_8-C_8. \label{VSVC}
\end{align}
which are straightforward to verify using the definitions Eqs (\ref{SO8chO})--(\ref{SO8chC}). They are relevant when we study superstring theories which have spacetime supersymmetry. In those cases the one-loop amplitudes would involve the overall factor $V_8-C_8$ or $V_8-S_8$, or their complex conjugates. Numerically they vanish, since we have $\theta[^1_1]=0$ so that $V_8-S_8=V_8-C_8\propto \theta[^0_0]^4-\theta[^0_1]^4-\theta[^1_0]^4$, which vanishes because of the identity
\begin{align}
	\theta[^0_0]^4-\theta[^0_1]^4-\theta[^1_0]^4 \equiv 0. \label{JID}
\end{align}
This results in a vanishing one-loop amplitude, accounting for the exact cancellation between the spacetime boson contribution and the spacetime fermion contribution. 

Despite the complexity in the construction of these compact notations, one can just keep in mind that they are power series of $q$ or $\b q$ with positive coefficients, and that the computation of one-loop amplitudes amounts largely to the manipulation of these characters and some left-over $\eta$'s and $\theta$'s.

\section{Unfolding the fundamental domain} \label{app2}

This appendix presents the scheme which we have used in Sec.\ref{TISG} for evaluating one-loop modular integrals on the fundamental domain of the group $SL(2,\mathbb Z)$: $\mfs F=\Big\{(\tau_1,\tau_2): -\demi < \tau_1 \leq \demi, \tau_2>\sqrt{1-\tau_1^2}\, \Big\}$. It is known as the unfolding trick or the orbit method, which can be applied to the integrations of the form
\begin{align}
	\ci=\int_{\mfs F} {d\tau_1 d\tau_2 \over \tau_2^2 } \sum_{m,n} f_{(m,n)}(\tau,\b \tau,\dots).\label{app2Eq1}
\end{align}
In the integrand the function $f_{(m,n)}(\tau,\b \tau,\dots)$ satisfies the following condition: setting $\tau'=(a\tau + b)/(c\tau +d)$ with $M=\left(\substack{a\ b \\c\ d}\right) \in SL(2,\mathbb Z)$, then we have $f_{(m,n)}(\tau',\b\tau',\dots)=f_{(m,n)M^{-1}}(\tau,\b \tau,\dots)$. It is then straightforward to verify the modular invariance of Eq.(\ref{app2Eq1}). The ellipses stand for spectator variables. The orbit method, as is demonstrated in \cite{McClain:1986id,O'Brien:1987pn,Bachas:1997mc}, asserts that the sum over $(m,n)$ can be decomposed into two orbits:
\begin{align}
	\ci=\int_{\mfs F} {d\tau_1 d\tau_2 \over \tau_2^2 }  f_{(0,0)}(\tau,\b \tau,\dots) +\int_{\sqcup} {d\tau_1d\tau_2 \over \tau_2^{2} } \sum_{m\neq 0} f_{(m,0)}(\tau,\b \tau,\dots), \label{UnFd2d}
\end{align}
where the $(0,0)$-orbit is still integrated on the fundamental domain and the $(m,0)$-orbit ($m\neq 0$) is integrated on the upper strip ($-\demi <\tau_1 \leq \demi$, $\tau_2>0$) denoted by $\sqcup$. 



The orbit method is efficient for a restrained class of situations, and it has the draw back of obscuring the symmetry of the integral. New techniques for evaluating the one-loop modular integrals are still developing. Recent progress can be found in \cite{IntFunAng}.

\end{document}